\documentclass[10pt, twoside]{article}

\usepackage[tbtags]{amsmath}
\usepackage{amsfonts}
\usepackage{amssymb}
\usepackage{latexsym}
\usepackage{mathrsfs}

\newcommand{\starMP}{*_{\scriptscriptstyle MC}}
\newcommand{\starM}{*_{\scriptscriptstyle M}}
\newcommand{\starD}{*_{\scriptscriptstyle D}}
\newcommand{\starP}{*_{\scriptscriptstyle C}}

\newcommand{\starEMC}{*_{\scriptscriptstyle EMC}}
\newcommand{\starSy}{*_{\scriptscriptstyle Sy}}

\newcommand{\cdotSy}{\cdot_{\scriptscriptstyle Sy}}
\newcommand{\cdotD}{\cdot_{\scriptscriptstyle D}}

\newcommand{\fett}[1]{\mbox{\boldmath$#1$}} 

\newcommand{\iu}{{\mathrm{i}}}

\begin{document}
\makeatletter
\title{Geometric Algebra and Star Products on the Phase Space}
\author{Peter Henselder\footnote{peter.henselder@uni-dortmund.de}\\
Fachbereich Physik, Universit\"at Dortmund\\
44221 Dortmund}

\maketitle

\begin{abstract}
Superanalysis can be deformed with a fermionic star product into a
Clifford calculus that is equivalent to geometric algebra. With
this multivector formalism it is then possible to formulate
Riemannian geometry and an inhomogeneous generalization of
exterior calculus. Moreover it is shown here how symplectic and
Poisson geometry fit in this context. The application of this
formalism together with the bosonic star product formalism of
deformation quantization leads then on space and space-time to a
natural appearance of spin structures and on phase space to BRST
structures that were found in the path integral formulation of
classical mechanics. Furthermore it will be shown that
Poincar\'{e} and Lie-Poisson reduction can be formulated in this
formalism.
\end{abstract}

\section{Introduction}
\qquad Geometric algebra was initiated by early ideas of Hamilton,
Grassmann and Clifford. The basic idea of geometrical algebra goes
back to Clifford, who combined the scalar and the wedge product of
vectors into one product in order to generalize complex analysis
to spaces of arbitrary dimensions. With this geometric or Clifford
product it is then possible to build up a multivector formalism
that contains the structures of vector analysis, complex analysis
and of spin. The description of spin was the physical motivation
to resume the program of Clifford calculus after the
Gibbs-Heavyside vector tuple formalism became the standard
formalism in physics. This was done by Hestenes
\cite{Hestenes1,Hestenes2} and independently also by K\"{a}hler
\cite{Kaehler}, who generalized the Clifford structures of Dirac
theory to an inhomogeneous exterior calculus and to curved spaces.
Since then geometric algebra was extended into a full formalism
and applied to a wide range of physical questions (see for example
\cite{Hestenes3,Doran1}).

In \cite{Doran2} it was noticed that geometric algebra can be
formulated in the realm of superanalysis, where a close connection
to pseudoclassical mechanics appeared. In the superanalytic
formulation of geometric algebra it is then possible to see that
the geometric product is a fermionic star product that deforms the
Grassmann structure into a Clifford structure \cite{Deform6}. Such
a fermionic star product appeared already in the founding paper of
deformation quantization \cite{Bayen1} and was applied in
\cite{Deform3,Deform5} for deformation quantization of
pseudoclassical mechanics and the formulation of spin and Dirac
theory in the star product formalism. In section two to six it
will be described following \cite{Doran1,Francis2} how geometric
algebra can be formulated as deformed superanalysis. The use of a
fermionic star product has the advantage that geometric algebra
can be unified with deformation quantization on a formal level.
The structure of the resulting formalism is supersymmetric. For
example one can describe rotations on the one hand with a bosonic
star exponential that acts on the bosonic coefficients of a
multivector and on the other hand with fermionic star
exponentials, i.e.\ rotors, that act on the fermionic basis
vectors of the multivector. A multivector is then invariant under
a combination of such rotations, which leads to a new
supersymmetric invariance condition. As shown in section seven the
condition of such an invariance on space and space-time leads to a
natural appearance of spin structures.

In section eight the superanalytic formulation of geometric
algebra will be applied to describe symplectic geometry and
Hamiltonian dynamics \cite{Hestenes9}. Using a superanalytic formalism 
leads to the question if the fermionic degrees of freedom on the phase 
space play a physical role just as the fermionic structures of space and
space-time constitute the spin. Moreover one can wonder if there
is a supersymmetry in classical mechanics. Supersymmetric
structures in classical mechanics were first noticed by Gozzi et
al.\ in the path integral description of classical mechanics
\cite{Gozzi1,Gozzi2,Gozzi3}. The classical path integral is a path
integral where all possible paths are constrained by a delta
function to the classical path. The delta function can be written
in terms of fermionic ghost degrees of freedom and the
corresponding Lagrange function that leads to the reduction to the
classical path as a gauge conditon was shown to be invariant under
a BRST- and an anti-BRST-transformation. Furthermore it was shown
that the fermionic ghosts could be interpreted as differential
forms on the phase space \cite{Gozzi4}. Together with the star
product formalism this was extended in \cite{Gozzi5} to a proposal
for a differential calculus in quantum mechanics. It will be shown
in section nine that these structures are the natural structures
of geometric algebra that appear if one considers bosonic and
fermionic time development on the phase space.

In the last two sections Poisson geometry and phase space
reduction will be discussed. It will be shown that geometric
algebra leads in this purely classical problem to a very elegant
formulation. Especially for the example of the rigid body one sees
that the dynamics is transferred by a fermionic rotor
transformation from the vector to the bivector level, which is the
same idea that is applied in the Kustanheimo-Stiefel
transformation \cite{Hestenes8}.

\section{Geometric Algebra and the Clifford Star Product}
\setcounter{equation}{0}\label{BFosc}
\qquad The Grassmann calculus of superanalysis can be deformed
with a fermionic star product into a Clifford calculus. This
Clifford calculus is a multivector calculus that is equivalent to
geometric algebra. In this superanalytic formulation of geometric
algebra the supernumbers correspond to the multivectors and the
fermionic star product corresponds to the geometric or Clifford
product. If one considers for example a $d$-dimensional vector
space with metric $\eta_{ij}$, the basis vectors of this vector
space are the Grassmann variables $\fett{\sigma}_i$,
$i=1,\ldots,d$. A vector $\fett{v}=v^i\fett{\sigma}_i$ is then a
supernumber of Grassmann grade one and a general multivector $A$
is a supernumber
\begin{equation}
A=A^0+A^i\fett{\sigma}_i+\frac{1}{2!}A^{i_1i_2}\fett{\sigma}_{i_1}
\fett{\sigma}_{i_2}+\ldots+\frac{1}{d!}A^{i_1\ldots i_d}
\fett{\sigma}_{i_1}\fett{\sigma}_{i_2}\ldots\fett{\sigma}_{i_d}.
\label{rbladebasis}
\end{equation}
The multivector $A$ is called $r$-vector if the highest appearing
Grassmann grade is $r$, i.e.\ if $A=\langle A\rangle_0+\langle
A\rangle_1+\ldots+\langle A\rangle_r$, where $\langle\;\rangle_n$
projects onto the term of Grassmann grade $n$. A multivector
$A_{(r)}$ is homogeneous or an $r$-blade if all appearing terms have the
same grade, i.e.\ if $A_{(r)}=\langle A_{(r)}\rangle_r$.

On the above vector space one can then define the Clifford star
product
\begin{equation}
A\starP B=A\,\exp\left[\sum_{i,j=1}^{d}\eta_{ij}
\frac{\overleftarrow{\partial}}{\partial\fett{\sigma}_i}\frac{\overrightarrow{\partial}}
{\partial\fett{\sigma}_j}\right]\,B.\label{externstarP}
\end{equation}
As a star product the Clifford star product acts in a distributive and
associative manner. The Clifford star product of two basis vectors is
$\fett{\sigma}_i\starP\fett{\sigma}_j=\eta_{ij}
+\fett{\sigma}_i\fett{\sigma}_j$ and with the metric one has
further $\fett{\sigma}_i\starP\fett{\sigma}^j=\delta_i^j
+\fett{\sigma}_i\fett{\sigma}^j$ and
$\fett{\sigma}^i\starP\fett{\sigma}^j=\eta^{ij}
+\fett{\sigma}^i\fett{\sigma}^j$. For two homogeneous multivectors
$A_{(r)}$ and $B_{(s)}$ the Clifford star product is the sum
\begin{equation}
A_{(r)}\starP B_{(s)}=\langle A_{(r)}\starP B_{(s)}\rangle_{r+s}
+\langle A_{(r)}\starP B_{(s)}\rangle_{r+s-2}+\cdots+\langle
A_{(r)}\starP B_{(s)}\rangle_{|r-s|}.
\end{equation}
The term of lowest and highest grade correspond to the inner and
the outer product
\begin{equation}
A_{(r)}\cdot B_{(s)}=\langle A_{(r)}\starP B_{(s)}\rangle_{|r-s|}
\qquad\mathrm{and}\qquad A_{(r)}B_{(s)}=\langle A_{(r)}\starP
B_{(s)}\rangle_{r+s}.
\end{equation}
Especially for the basis vectors one has
$\fett{\sigma}_i\cdot\fett{\sigma}_j=\eta_{ij}$,
$\fett{\sigma}_i\cdot\fett{\sigma}^j=\delta_i^j$ and
$\fett{\sigma}^i\cdot\fett{\sigma}^j=\eta^{ij}$.

As a first example one can consider the two dimensional euclidian
case. A general element of the Clifford algebra is a supernumber
$A=A^0+A^1\fett{\sigma}_1+A^2\fett{\sigma}_2
+A^{12}\fett{\sigma}_1\fett{\sigma}_2$ and a vector corresponds to
the supernumber $\fett{a}=a^1\fett{\sigma}_1+a^2\fett{\sigma}_2$.
The Clifford star product of two vectors is
\begin{equation}
\fett{a}\starP\fett{b}=\fett{ab}+\fett{a}\left[\sum_{n=1}^2
\frac{\overleftarrow{\partial}}{\partial\fett{\sigma}_n}
\frac{\overrightarrow{\partial}}{\partial\fett{\sigma}_n}\right]\fett{b}
=(a^1b^2-a^2b^1)\fett{\sigma}_1\fett{\sigma}_2+a^1b^1+a^2b^2\equiv
\fett{a}\wedge\fett{b}+\fett{a}\cdot\fett{b},
\end{equation}
The basis bivector $I_{(2)}=\fett{\sigma}_1\fett{\sigma}_2$
fulfills $I_{(2)}\starP I_{(2)}=I^{2\starP}_{(2)}=-1$ and
$\overline{I_{(2)}}=-I_{(2)}$, where the involution reverses the
order of the Grassmann elements:
\begin{equation}
\overline{\fett{\sigma}_{i_1}\ldots\fett{\sigma}_{i_r}}
=\fett{\sigma}_{i_r}\ldots\fett{\sigma}_{i_1}.
\end{equation}
$I_{(2)}$ plays in two dimensions the role of a imaginary unit.

In three dimensions a general Clifford number can be written as
\begin{equation}
A=A^0+a^1\fett{\sigma}_1+A^2\fett{\sigma}_2+A^3\fett{\sigma}_3
+A^{12}\fett{\sigma}_1\fett{\sigma}_2 +A^{13}\fett{\sigma}_3\fett{\sigma}_1
+A^{23}\fett{\sigma}_2\fett{\sigma}_3 +A^{123}\fett{\sigma}_1\fett{\sigma}_2
\fett{\sigma}_3 \label{A3D}
\end{equation}
and it contains a scalar, a vector, a bivector and a trivector or
pseudoscalar part. The basis bivectors
$\mathtt{Q}_1=\fett{\sigma}_2\fett{\sigma}_3$,
$\mathtt{Q}_2=\fett{\sigma}_1\fett{\sigma}_3$ and
$\mathtt{Q}_3=\fett{\sigma}_1\fett{\sigma}_2$ fulfill
\begin{equation}
\mathtt{Q}_1^{2\starP}=\mathtt{Q}_2^{2\starP}=\mathtt{Q}_3^{2\starP}
=\mathtt{Q}_1\starP\mathtt{Q}_2\starP\mathtt{Q}_3=-1
\end{equation}
which means that the even multivector $Q=q^0+q^i\mathtt{Q}_i$ is a
quaternion. The trivector part with basis
$I_{(3)}=\fett{\sigma}_1\fett{\sigma}_2\fett{\sigma}_3$ can be
used to describe the duality of vectors
$\fett{b}=b^1\fett{\sigma}_1+b^2\fett{\sigma}_2
+b^3\fett{\sigma}_3$ and bivectors
$\mathtt{B}=b^1\fett{\sigma}_2\fett{\sigma}_3+b^2\fett{\sigma}_3\fett{\sigma}_1
+b^3\fett{\sigma}_1\fett{\sigma}_2$, i.e.\
$\mathtt{B}=I_{(3)}\starP\fett{b}$. With this relation one can
then write the geometric product of two vectors
$\fett{a}=a^1\fett{\sigma}_1+a^2 \fett{\sigma}_2
+a^3\fett{\sigma}_3$ and $\fett{b}=b^1\fett{\sigma}_1+b^2
\fett{\sigma}_2+b^3\fett{\sigma}_3$ as:
\begin{equation}
\fett{a}\starP \fett{b}=\fett{a}\cdot \fett{b}
+I_3\starP(\fett{a}\times \fett{b}),\label{astarPb}
\end{equation}
where $\fett{a}\cdot\fett{b}=\sum_{k=1}^{3}a^kb^k$ and
$\fett{a}\times\fett{b}
=\varepsilon_{kl}^{\hphantom{kl}m}a^kb^l\fett{\sigma}_m$.

\section{Vector Manifolds}
In geometric algebra the points of a manifold are treated as
vectors, so that a manifold can be seen as a surface in a flat
ambient space. The at least $(d+1)$-dimensional flat ambient space
is spanned by the rectangular basis vectors $\fett{\sigma}_a$ and
is equipped with the constant metric $\eta_{ab}$. A
$d$-dimensional vector manifold with coordinates $x^i$,
$i=1,\ldots,d$ that is embedded in this ambient vector space is
then described by smooth functions $f^a(x^i)$ and has the form
$\fett{x}(x^i) =f^a(x^i)\fett{\sigma}_a$, one also uses the
notation $\fett{x}(x^i) =x^a(x^i)\fett{\sigma}_a$. The vectors
\begin{equation}
\fett{\xi}_i(\fett{x})=\frac{\partial\fett{x}}{\partial x^i}
\label{framexi}
\end{equation}
are the frame vectors of the manifold, which can be expanded in
the ambient space as $\fett{\xi}_i(\fett{x})=\xi^a_i(\fett{x})
\fett{\sigma}_a$. The $\fett{\xi}_i(\fett{x})$ span the tangent
space $T_{\mbox{\scriptsize \boldmath$x$}}M$, with the
Clifford star product
\begin{equation}
F\starP G=F\,\exp\left[\sum_{i,j=1}^d g_{ij}(\fett{x})
\frac{\overleftarrow{\partial}}{\partial\fett{\xi}_i}
\frac{\overrightarrow{\partial}}{\partial\fett{\xi}_j}\right]\,G.
\label{internstarP}
\end{equation}
The scalar product of two frame vectors can be calculated
internally and externally as $g_{ij}=\fett{\xi}_i\cdot
\fett{\xi}_j=(\xi^a_i\fett{\sigma}_a)\cdot
(\xi^b_j\fett{\sigma}_b)$, so that $\xi^a_i\xi^b_j\eta_{ab}
=g_{ij}$. In general one has for both, the Clifford star product
of the ambient space and the intrinsic Clifford star product
(\ref{internstarP}):
\begin{equation}
\fett{\xi}_i\starP\fett{\xi}_j=g_{ij}+\fett{\xi}_i\fett{\xi}_j,\qquad
\fett{\xi}_i\starP\fett{\xi}^j=\delta_i^j+\fett{\xi}_i\fett{\xi}^j,
\qquad\mathrm{and}\qquad
\fett{\xi}^i\starP\fett{\xi}^j=g^{ij}+\fett{\xi}^i\fett{\xi}^j.
\end{equation}

For an orientable manifold there exists a global unit-pseudoscalar
$I_{(d)}(\fett{x})=\fett{\xi}_1\fett{\xi}_2\ldots\fett{\xi}_d/
|\fett{\xi}_1\fett{\xi}_2\ldots\fett{\xi}_d|$, which allows to
define a projector $P$ on the vector manifold that projects an
arbitrary multivector $A(\fett{x})$ in the ambient space onto the
vector manifold:
\begin{equation}
P(A(\fett{x}),\fett{x})=(A(\fett{x})\cdot I_{(d)}(\fett{x}))\starP
I^{-1\starP}_{(d)}(\fett{x}).\label{PA}
\end{equation}
A vector $\fett{v}(\fett{x})=v^a(x^i)\fett{\sigma}_a$ in a point
of the vector manifold can then be decomposed into an intrinsic
part $P(\fett{v})=(\fett{\xi}_i\cdot\fett{v})\fett{\xi}^i
=(v_a\xi^a_i)\fett{\xi}^i$ which is tangent to the manifold and an
extrinsic part $P_{\bot}(\fett{v})=\fett{v}-P(\fett{v})$. Applying
the projector to the nabla operator of the ambient space gives a
vector derivative intrinsic to the manifold:
\begin{equation}
\fett{\partial}=P(\fett{\nabla})=\fett{\xi}^i(\fett{\xi}_i\cdot
\fett{\nabla})=\fett{\xi}^i(\xi^a_i\partial_a)=\fett{\xi}^i\partial_i
\label{fettpartidef}
\end{equation}
and for a tangent vector $\fett{a}$ the directional derivative in
the $\fett{a}$-direction is
$\fett{a}\cdot\fett{\partial}=a^i\partial_i=a^i\xi^a_i\partial_a
=\fett{a}\cdot\fett{\nabla}$. With the intrinsic vector derivative
(\ref{fettpartidef}) the cotangent frame vectors $\fett{\xi}^i$
can also be obtained as the gradient of the coordinate functions
$x^i(\fett{x})$ that arise from the inversion of the vector
manifold parametrization $\fett{x}=\fett{x}(x^i)$:
\begin{equation}
\fett{\xi}^i=\fett{\partial}x^i.\label{diffo}
\end{equation}

If one now applies the directional derivative
$\fett{a}\cdot\fett{\partial}$ on a tangent multivector field
$A(\fett{x})$ the result does not in general lie completely inside
the manifold. So if one wants to have a purely intrinsic result
one has to use the projection operator $P$ again. This leads to
the definition of a new type of derivative that acts on tangent
multivector fields and returns tangent multivector fields. This
new derivative is the covariant derivative and is defined by:
\begin{equation}
(\fett{a}\cdot\fett{D})A(\fett{x})
=P\big((\fett{a}\cdot\fett{\partial})A(\fett{x})\big).
\end{equation}
In the case of a scalar field $f(\fett{x})$ on the manifold the
covariant and the intrinsic derivative are the same:
\begin{equation}
(\fett{a}\cdot\fett{\partial})f =(\fett{a}\cdot\fett{D})f,
\label{partiska}
\end{equation}
while for tangent vector fields $\fett{a}$ and $\fett{b}$ one has
\begin{equation}
(\fett{a}\cdot\fett{\partial})\fett{b}=P\big((\fett{a}\cdot\fett{\partial})
\fett{b}\big)+P_{\bot}\big((\fett{a}\cdot\fett{\partial})\fett{b}\big)
=(\fett{a}\cdot\fett{D})\fett{b}+\fett{b}\cdot
\mathtt{S}(\fett{a}), \label{partivec}
\end{equation}
where $\mathtt{S}(\fett{a})$ is the so called shape tensor, which
is a bivector that describes both intrinsic and extrinsic
properties of the vector manifold. The multivector generalization
of (\ref{partivec}) is
\begin{equation}
(\fett{a}\cdot\fett{\partial})A=(\fett{a}\cdot\fett{D})A+A\times
\mathtt{S}(\fett{a}),\label{partimultvec}
\end{equation}
where $A\times B=\frac{1}{2}(A\starP B-B\starP A)
=\frac{1}{2}\left[A,B\right]_{\starP}$ is the commutator product
(not to be confused with the vector cross product used in
(\ref{astarPb}); the cross product of two three-dimensional
vectors $\fett{a}$ and $\fett{b}$ and the commutator product of
the corresponding bivectors $\mathtt{A}=I_{(3)}\starP\fett{a}$ and
$\mathtt{B}=I_{(3)}\starP\fett{b}$ are connected according to
$-I_{(3)}\starP(\fett{a}\times\fett{b}) =\frac{1}{2}
\left[I_{(3)}\starP\fett{a},I_{(3)}\starP\fett{b}\right]_{\starP}
=\mathtt{A}\times\mathtt{B}$ ). The commutator product of an
$r$-vector and a bivector gives again an $r$-vector so that all
terms in (\ref{partimultvec}) are $r$-vectors. Furthermore it is
clear that (\ref{partimultvec}) reduces to (\ref{partivec}) if $A$
is a vector field and to (\ref{partiska}) if $A$ is a scalar
field.

As a tangent vector $(\fett{a}\cdot\fett{D})\fett{b}$ can be
expanded in the $\fett{\xi}_i$ base:
\begin{equation}
(\fett{a}\cdot\fett{D})\fett{b}=a^j\big((D_jb^i)\fett{\xi}_i
+b^i(D_j\fett{\xi}_i)^k\fett{\xi}_k\big)
=a^j\big(\partial_jb^i+b^k\Gamma^i_{jk}\big)\fett{\xi}_i,
\label{aDbcoeff}
\end{equation}
where $\Gamma_{jk}^i=(D_j\fett{\xi}_k)\cdot\fett{\xi}^i
=\big(D_j\fett{\xi}_k\big)^i$ is the $i$-th component of
$D_j\fett{\xi}_k$, which extrinsically can be written as
$\Gamma_{jk}^i=\big(D_j\xi^a_k\fett{\sigma}_a\big)\cdot\xi^i_b\fett{\sigma}^b
=(\partial_j\xi_k^a)\xi_a^i$. One of the properties the
$\Gamma_{ij}^k$ fulfill is the metric compatibility
$\partial_kg_{ij}-\Gamma_{ki}^lg_{lj}-\Gamma_{kj}^lg_{li}=0$, which
can be found if one applies $D_k$ on both sides of
$\fett{\xi}_i\cdot\fett{\xi}_j=g_{ij}$. This means that the
$\Gamma_{jk}^i$ are the Christoffel symbols and
$(\fett{a}\cdot\fett{D})\fett{b}$ is the Levi-Civita connection.
The symmetry of the $\Gamma_{jk}^i$ in the lower indices is a
consequence of
\begin{equation}
\partial_i\fett{\xi}_j-\partial_j\fett{\xi}_i=(\partial_i\partial_j
-\partial_j\partial_i)\fett{x}=0.\label{particommux}
\end{equation}
Projecting into the manifold gives
$D_i\fett{\xi}_j-D_j\fett{\xi}_i=0$, so that the symmetry of the
$\Gamma_{jk}^i$ in the lower indices follows. From
(\ref{particommux}) follows further, that
\begin{equation}
(\fett{a}\cdot\fett{\partial})\fett{b}
-(\fett{b}\cdot\fett{\partial})\fett{a} =\big(a^j(\partial_jb^i)
-b^j(\partial_ja^i)\big)\fett{\xi}_i \label{Liederiveccompo}
\end{equation}
is an intrinsic quantity that corresponds to the Lie derivative or
the Jacobi-Lie bracket
\begin{equation}
\mathscr{L}_{\mbox{\scriptsize{\boldmath$a$}}}\fett{b}
=\left[\fett{a},\fett{b}\right]_{JLB} \equiv
(\fett{a}\cdot\fett{\partial})\fett{b}
-(\fett{b}\cdot\fett{\partial})\fett{a}
=(\fett{a}\cdot\fett{D})\fett{b} -(\fett{b}\cdot\fett{D})\fett{a}.
\label{Liederivec}
\end{equation}
The holonomy condition (\ref{particommux}) can then be written
with $\fett{\xi}_i\cdot\fett{\partial} =\partial_i$ in the more
familiar form $\mathscr{L}_{\mbox{\scriptsize{\boldmath$\xi$}}_i}
\fett{\xi}_j=\left[\fett{\xi}_i,\fett{\xi}_j\right]_{JLB}=0$. One
can also conclude with (\ref{partivec}) that since
$\left[\fett{a},\fett{b}\right]_{JLB}$ is an intrinsic quantity,
the extrinsic parts in the Jacobi-Lie bracket have to cancel, i.e.
$\fett{a}\cdot \mathtt{S}(\fett{b})
=\fett{b}\cdot\mathtt{S}(\fett{a})$.

A natural generalization of the Lie derivative to multivectors is
given by the Schouten-Nijenhuis bracket
\begin{equation}
\mathscr{L}_{A_{(r)}}B_{(s)}=\left[A_{(r)},B_{(s)}\right]_{SNB}
=(-1)^{r-1}(A_{(r)}\cdot\fett{D})B_{(s)}
+(-1)^{rs}(-1)^{s-1}(B_{(s)}\cdot\fett{D})A_{(r)}. \label{SNBDef}
\end{equation}
The Schouten-Nijenhuis bracket can be written in this way due
to the fact that (\ref{SNBDef}) has the grade $r+s-1$,
fulfills
\begin{eqnarray}
\left[A_{(r)},B_{(s)}\right]_{SNB}&=&(-1)^{rs}\left[B_{(s)},A_{(r)}\right]_{SNB},\\
\left[A_{(r)},B_{(s)}C_{(t)}\right]_{SNB}
&=&\left[A_{(r)},B_{(s)}\right]_{SNB}C_{(t)}+(-1)^{rs+s}B_{(s)}
\left[A_{(r)},C_{(t)}\right]_{SNB}
\end{eqnarray}
and reduces for scalar functions $f$, $g$ and vector fields
$\fett{a}$ and $\fett{b}$ to
\begin{equation}
\left[f,g\right]_{SNB}=0,\qquad
\left[\fett{a},f\right]_{SNB}=(\fett{a}\cdot\fett{D})f
\qquad\mathrm{and}\qquad \left[\fett{a},\fett{b}\right]_{SNB}
=\mathscr{L}_{\mbox{\scriptsize{\boldmath$a$}}}\fett{b}.
\end{equation}
Furthermore one has the super-Jacobi-identity
\begin{multline}
(-1)^{rt}\left[\left[A_{(r)},B_{(s)}\right]_{SNB},C_{(t)}\right]_{SNB}
+(-1)^{rs}\left[\left[B_{(s)},C_{(t)}\right]_{SNB},A_{(r)}\right]_{SNB}\\
+(-1)^{st}\left[\left[C_{(t)},A_{(r)}\right]_{SNB},B_{(s)}\right]_{SNB}
=0.
\end{multline}

\section{Exterior Calculus}
The exterior calculus \cite{Francis2} can be constructed by noting
that the cotangent frame vector or one-form (\ref{diffo}) can be
written with (\ref{partiska}) as
\begin{equation}
\fett{\xi}^i=\fett{D}x^i=\fett{\partial}x^i\equiv\fett{d}x^i.
\label{xiio}
\end{equation}
In order to see how the directional covariant derivative acts on a
general one-form $\fett{\omega}=\omega_i\fett{\xi}^i$ one first
applies $D_j$ on both sides of
$\fett{\xi}^i\cdot\fett{\xi}_k=\delta^i_k$ which gives
$(D_j\fett{\xi}^i)\cdot\fett{\xi}_k=(D_j\fett{\xi}^i)_k=-\Gamma_{jk}^i$,
so that the covariant derivative of $\fett{\omega}$ reads
\begin{equation}
(\fett{a}\cdot\fett{D})\fett{\omega}=a^j\big((D_j\omega_i)\fett{\xi}^i
+\omega_i(D_j\fett{\xi}^i)_k\fett{\xi}^k\big)=a^j\big(\partial_j\omega_i
-\omega_k\Gamma_{ji}^k\big)\fett{\xi}^i.
\end{equation}
Furthermore it is easy to see that $\fett{dd}x^i=\fett{D\xi}^i=0$. The 
closedness of $\fett{\xi}^i$ can be used to
calculate the relation of the $\Gamma_{jk}^i$ and the metric:
\begin{eqnarray}
\Gamma_{jk}^i=(D_j\fett{\xi}_k)\cdot\fett{\xi}^i&=&\frac{1}{2}
\left[(D_j\fett{\xi}_k)+(D_k\fett{\xi}_j)\right]\cdot\fett{\xi}^i \label{Gamma1}\\
&=&\frac{1}{2}\left[\fett{\xi}_j\cdot(\fett{D}\fett{\xi}_k)+
\Gamma_{mk}^lg_{jl}\fett{\xi}^m+\fett{\xi}_k\cdot(\fett{D}\fett{\xi}_j)
+\Gamma_{mj}^lg_{kl}\fett{\xi}^m\right]\cdot\fett{\xi}^i\label{Gamma2}\\
&=&\frac{1}{2}\left[\fett{\xi}_j\cdot(\fett{D}g_{km}\fett{\xi}^m)
+\fett{\xi}_k\cdot(\fett{D}g_{jm}\fett{\xi}^m)+(\partial_mg_{jk})\fett{\xi}^m
\right]\cdot\fett{\xi}^i\label{Gamma3}\\
&=&\frac{1}{2}\left[(\partial_ng_{km})\fett{\xi}_j\cdot
\fett{\xi}^n\fett{\xi}^m+(\partial_ng_{jm})\fett{\xi}_k\cdot
\fett{\xi}^n\fett{\xi}^m+(\partial_mg_{jk})\fett{\xi}^m\right]
\cdot\fett{\xi}^i\label{Gamma4}\\
&=&\frac{1}{2}\left[(\partial_ng_{km})(\delta_j^n\fett{\xi}^m
-\delta_j^m\fett{\xi}^n)+(\partial_ng_{jm})(\delta_k^n\fett{\xi}^m
-\delta_k^m\fett{\xi}^n)+(\partial_mg_{jk})\fett{\xi}^m\right]\cdot\fett{\xi}^i\\
&=&\frac{1}{2}g^{il}\left[\partial_jg_{kl}+\partial_kg_{jl}
-\partial_lg_{jk}\right],\label{Gammaijkgij}
\end{eqnarray}
where one uses in (\ref{Gamma1})
\begin{equation}
\fett{\xi}_j\cdot (\fett{D}\fett{\xi}_k)=\fett{\xi}_j\cdot
(\fett{\xi}^i D_i\fett{\xi}_k)=\fett{\xi}_j\cdot
(\fett{\xi}^i\Gamma_{ik}^l\fett{\xi}_l)
=\Gamma_{ik}^l(\delta_j^i\fett{\xi}_l-g_{jl}\fett{\xi}^i)
=D_j\fett{\xi}_k-\Gamma_{ik}^lg_{jl}\fett{\xi}^i.
\end{equation}

With expression (\ref{Gammaijkgij}) it is possible to show that
the shape bivector can be written as
\begin{equation}
\mathtt{S}(\fett{a})=\frac{1}{2}\big(\fett{\xi}^i\fett{\partial}a_i
-\fett{\xi}_i\fett{\partial}a^i+\fett{\xi}^i(\fett{a}\cdot
\fett{\partial})\fett{\xi}_i\big),
\end{equation}
or $\mathtt{S}_i=\mathtt{S}(\fett{\xi}_i)=\frac{1}{2}
\fett{\xi}^j\fett{\xi}^k\partial_k g_{ij}
+\frac{1}{2}\fett{\xi}^j\partial_i\fett{\xi}_j$. This can be
proved by calculating
\begin{eqnarray}
\!\!\!\!\!\!\!\!\!\!\!\!\!\!\!\!\!\fett{b}\cdot
\mathtt{S}(\fett{a})&=&
\frac{1}{2}\big(b^i(\partial_ja_i)\fett{\xi}^j-b^j(\partial_ja_i)
\fett{\xi}^i-b_i(\partial_ja^i)\fett{\xi}^j+b^j(\partial_ja^i)\fett{\xi}_i
+a^jb^i(\partial_j\fett{\xi}_i)-a^jb^k\fett{\xi}^i(\fett{\xi}_k\cdot
\partial_j\fett{\xi}_i)\big)\\
&=&\frac{1}{2}\big(a^kb^i(\partial_jg_{ik})\fett{\xi}^j-a^kb^i(\partial_i
g_{jk})\fett{\xi}^j+a^ib^j(\partial_i\fett{\xi}_j)-a^jb^k\fett{\xi}^i
(\fett{\xi}_k\cdot\partial_j\fett{\xi}_i)\big)\label{bSa2}\\
&=&-a^kb^i\Gamma_{ki}^l\fett{\xi}_l+\frac{1}{2}a^jb^k\fett{\xi}^i
(\partial g_{ki})+\frac{1}{2}a^ib^j(\partial_i\fett{\xi}_j)
-\frac{1}{2}a^jb^k\fett{\xi}^i(\fett{\xi}_k\cdot\partial_j\fett{\xi}_i)\label{bSa3}\\
&=&-a^kb^i\Gamma_{ki}^l\fett{\xi}_l+\frac{1}{2}a^ib^j\fett{\xi}^k\left[
(\partial_i\fett{\xi}_j)\cdot\fett{\xi}_k\right]+\frac{1}{2}a^ib^j
(\partial_i\fett{\xi}_j)\label{bSa4}\\
&=&a^ib^j(\partial_i\fett{\xi}_j)-a^ib^j\Gamma_{ij}^k\fett{\xi}_k\label{bSa5}\\
&=&(\fett{a}\cdot\fett{\partial})\fett{b}-(\fett{a}\cdot\fett{D})\fett{b}\\
&=&P_{\bot}\big((\fett{a}\cdot\fett{\partial})\fett{b}\big),
\end{eqnarray}
which corresponds to definition (\ref{partivec}). In
(\ref{bSa2}) relation (\ref{Gammaijkgij}) was used and in
(\ref{bSa4}) one uses
\begin{equation}
\fett{\xi}^k\left[(\partial_i\fett{\xi}_j)\cdot\fett{\xi}_k\right]
=\xi^k_c\fett{\sigma}^c\left[(\partial_i\xi_j^a\fett{\sigma}_a)
\cdot\xi^b_k\fett{\sigma}_b\right]
=(\partial_i\xi_j^a)\fett{\sigma}_a=\partial_i\fett{\xi}_j.
\end{equation}

While the exterior derivative of the reciprocal basis vectors is
zero, the exterior derivative of a general one-form
$\fett{\omega}=\omega_i\fett{\xi}^i$ is a two-form
$\fett{d\omega}=(\fett{D}\omega_j)\fett{\xi}^j
+\omega_j\fett{D}\fett{\xi}^j
=(\partial_i\omega_j)\fett{\xi}^i\fett{\xi}^j$. A general $r$-form
is then a covariant $r$-blade $A^{(r)}$ \cite{Francis2} and can be
written as
\begin{equation}
A^{(r)}=\frac{1}{r!}A_{i_1i_2\ldots i_r}\fett{d}x^{i_1}
\fett{d}x^{i_2}\ldots\fett{d}x^{i_r} =\frac{1}{r!}A_{i_1i_2\ldots
i_r}\fett{\xi}^{i_1}\fett{\xi}^{i_2} \ldots\fett{\xi}^{i_r}.
\end{equation}
Applying the exterior differential, to $A^{(r)}$ gives 
\begin{equation}
\fett{d}A^{(r)}=\frac{1}{r!}\left(\frac{\partial A_{i_1i_2\ldots
i_r}} {\partial
x^j}\right)\fett{d}x^j\fett{d}x^{i_1}\fett{d}x^{i_2}\ldots
\fett{d}x^{i_r}=\frac{1}{r!}\left(\frac{\partial A_{i_1i_2\ldots
i_r}} {\partial
x^j}\right)\fett{\xi}^j\fett{\xi}^{i_1}\fett{\xi}^{i_2}\ldots
\fett{\xi}^{i_r},
\end{equation}
which is a $(r+1)$-form or a covariant $(r+1)$-blade.

It is then also straight forward to translate other structures of
exterior calculus into the language of superanalytic geometric
algebra, for example the Hodge dual is given by
\begin{equation}
\star\left(\fett{\xi}^{i_1}\fett{\xi}^{i_2}\ldots\fett{\xi}^{i_r}\right)
=\frac{\sqrt{|g|}}{(d-r)!}\varepsilon^{i_1\ldots i_r}
_{\hphantom{i_1\ldots i_r}i_{r+1}\ldots
i_d}\fett{\xi}^{i_{r+1}}\ldots \fett{\xi}^{i_d},\label{Hodgedef}
\end{equation}
with $\varepsilon^{i_1\ldots i_r} _{\hphantom{i_1\ldots
i_r}i_{r+1}\ldots i_d}=g^{i_1j_1}\ldots g^{i_rj_r}
\varepsilon_{j_1\ldots j_ri_{r+1}\ldots i_d}$ and
$\varepsilon_{i_1\ldots i_d}=1$ for even permutations. In the
euclidian or Minkowski case the Hodge dual can be written as
\begin{equation}
\star A^{(r)}=(-1)^{(d-r)r+r(r-1)/2}I^{(d)}\starP A^{(r)}
\end{equation}
and the inverse Hodge star operator in the euclidian case as
\begin{equation}
\star^{-1}A^{(r)}=(-1)^{r(d-r)}\star A^{(r)}
=(-1)^{(r-1)r/2}I^{(d)}\starP A^{(r)},
\end{equation}
while in the four dimensional Minkowski case one has an additional
minus sign, i.e.\ $\star^{-1}=(-1)^{r(d-r)+1}\star$. With the
Hodge star operator as defined in (\ref{Hodgedef}) the
coderivative $\fett{d}^{\dagger}$ is given in the Riemannian case
as
\begin{equation}
\fett{d}^{\dagger} A^{(r)}=(-1)^{dr+d+1}\star\fett{d}\star A^{(r)}
\end{equation}
and in the Minkowski case as $\fett{d}^{\dagger}
A^{(r)}=(-1)^{dr+d}\star\fett{d}\star A^{(r)}$. Writing this down
in components one sees that the coderivative maps an $r$-form into 
an $(r-1)$-form and can be written
as $\fett{d}^{\dagger} A^{(r)}=-\fett{d}\cdot A^{(r)}$.

The interior product that maps an $r$-blade $A^{(r)}$ into an
$(r-1)$-blade is just the scalar product with a vector, which can
be generalized to the case of two multivectors $A_{(r)}$ and
$B^{(s)}$ as
\begin{equation}
\dot{\iota}_{A_{(r)}}B^{(s)}=\overline{A_{(r)}}\cdot B^{(s)},
\label{iArBs}
\end{equation}
so that one has for example
\begin{eqnarray}
\left(\overline{\fett{a}_1\fett{a}_2\ldots\fett{a}_{r+1}}\right)\cdot
\fett{d}A^{(r)}
&=&\sum_{n=1}^{r+1}(-1)^{n+1}(\fett{a}_n\cdot\fett{\partial})
\left(\overline{\fett{a}_1\ldots\check{\fett{a}}_n\ldots\fett{a}_{r+1}}\right)
\cdot A^{(r)}\nonumber\\
&&+\sum_{m<n}(-1)^{m+n}\left(\overline{\left[\fett{a}_m,\fett{a}_n\right]_{JLB}
\fett{a}_1\ldots\check{\fett{a}}_m\ldots\check{\fett{a}}_n
\ldots\fett{a}_{r+1}}\right)\cdot A^{(r)}.\label{iAr}
\end{eqnarray}
In the same way Cartan's magic formula
\begin{equation}
\mathscr{L}_{\mbox{\scriptsize \boldmath$a$}}\fett{\omega}
=\left(\fett{d}\dot{\iota}_{\mbox{\scriptsize \boldmath$a$}}
+\dot{\iota}_{\mbox{\scriptsize \boldmath$a$}} \fett{d}\right)
\fett{\omega}=\left(a^i(\partial_i\omega_j)+(\partial_ja^i)
\omega_i\right)\fett{\xi}^j,\label{magicCartan}
\end{equation}
generalizes to
\begin{equation}
\mathscr{L}_{\mbox{\scriptsize \boldmath$a$}}A^{(r)}
=\left(\fett{d}\dot{\iota}_{\mbox{\scriptsize \boldmath$a$}}
+\dot{\iota}_{\mbox{\scriptsize \boldmath$a$}}
\fett{d}\right)A^{(r)}=\fett{D}(\fett{a}\cdot A^{(r)})
+\fett{a}\cdot(\fett{D}A^{(r)}).\label{Liederiform}
\end{equation}

Up to now only the coordinate basis of the $\fett{\xi}_i$ was
used, in general it is also possible to use a non-coordinate basis
given by
\begin{equation}
\fett{\vartheta}_r=\vartheta_r^i\fett{\xi}_i\qquad\mathrm{and}\qquad
\fett{\xi}_i=\vartheta_i^r\fett{\vartheta}_r,
\end{equation}
where $\vartheta_r^i=\fett{\vartheta}_r\cdot\fett{\xi}^i$ are
functions of the $x^k$, with
$\vartheta_i^r\vartheta_r^j=\delta_i^j$ and $g_{ij}=\vartheta_i^r
\vartheta_j^sg_{rs}$. Analogously the reciprocal non-coordinate
basis $\fett{\vartheta}^r$ can be expanded with the
$\vartheta_i^r$ in the reciprocal coordinate basis of the
$\fett{\xi}^i$. A special choice for the non-coordinate frame
fields is obtained by the conditions
$\fett{\vartheta}_r\cdot\fett{\vartheta}_s =\eta_{rs}$ and
$\partial_i\fett{\vartheta}_r=0$. This means the
$\fett{\vartheta}_r$ span a (pseudo)-euclidian base and they move
on the vector manifold so that
\begin{equation}
D_i\fett{\vartheta}_r=-\fett{\vartheta}_r\cdot\mathtt{S}_i.
\end{equation}
This shows that the shape tensor, which has in the
$\fett{\vartheta}_r$-frame the form $\mathtt{S}_r=
\mathtt{S}(\fett{\vartheta}_r)=\vartheta_r^i\mathtt{S}_i$, is
proportional to the Fock-Ivanenko bivector $\Gamma_i$ \cite{Fock},
i.e.\ $\mathtt{S}_i=-2\Gamma_i$.

For general non-coordinate basis vectors the Jacobi-Lie bracket is
no longer zero, one rather has
\begin{eqnarray}
\left[\fett{\vartheta}_r,\fett{\vartheta}_s\right]_{JLB}&=&
\vartheta_r^i(\fett{\xi}_i\cdot\fett{D})(\vartheta_s^j\fett{\xi}_j)
-\vartheta_s^i(\fett{\xi}_i\cdot\fett{D})(\vartheta_r^j\fett{\xi}_j)\\
&=&\vartheta_r^i\left[(D_i\vartheta_s^j)\fett{\xi}_j
+\vartheta_s^j(D_i\fett{\xi}_j)\right]
-\vartheta_s^i\left[(D_i\vartheta_r^j)\fett{\xi}_j
+\vartheta_r^j(D_i\fett{\xi}_j)\right]\\
&=&\left[\vartheta_r^iD_i\vartheta_s^j
-\vartheta_s^iD_i\vartheta_r^j\right]\fett{\xi}_j\\
&=&\left[\partial_r\vartheta_s^j
-\partial_s\vartheta_r^j\right]\vartheta_j^t
\fett{\vartheta}_t\\
&=&C_{rs}^t\fett{\vartheta}_t,
\end{eqnarray}
with
$C_{rs}^t=\left[\fett{\vartheta}_r,\fett{\vartheta}_s\right]_{JLB}\cdot
\fett{\vartheta}^t=\left[\partial_r\vartheta_s^j
-\partial_s\vartheta_r^j\right]\vartheta_j^t$. For tangent vector
fields $\fett{a}=a^r\fett{\vartheta}_r$ and
$\fett{b}=b^s\fett{\vartheta}_s$, it follows then that
\begin{equation}
\mathscr{L}_{\mbox{\scriptsize{\boldmath$a$}}}\fett{b}=\left[\fett{a},
\fett{b}\right]_{JLB}=\big(a^r(\partial_rb^s)-b^r(\partial_ra^s)\big)
\fett{\vartheta}_s+a^rb^s\left[\fett{\vartheta}_r,\fett{\vartheta}_s
\right]_{JLB},
\end{equation}
which reduces in a coordinate basis to (\ref{Liederiveccompo}).

In the non-coordinate basis a straight forward calculation shows
that the $\Gamma_{rs}^t$ are given by
\begin{equation}
\Gamma_{rs}^t=-\left[(\fett{\vartheta}_r\cdot\fett{D})
\fett{\vartheta}^t\right]\cdot\fett{\vartheta}_s
=\frac{1}{2}g^{tu}\left[\partial_rg_{su}+\partial_sg_{ru}
-\partial_ug_{rs}\right]+\frac{1}{2}g^{tu}(C_{urs}+C_{usr}-C_{sru}),
\end{equation}
where $C_{rsu}=g_{tu}C^t_{rs}$. While in the coordinate base
$\left[\fett{\xi}_i,\fett{\xi}_j\right]_{JLB}=0$ insured that the
$\Gamma_{ij}^k$ are symmetric in the lower indices, one has in the
non-coordinate basis the relation
$\Gamma_{rs}^t-\Gamma_{sr}^t=C_{rs}^t$. This implies that the
non-coordinate one-forms $\fett{\vartheta}^r$ are not closed:
\begin{eqnarray}
\fett{d}\fett{\vartheta}^r=\fett{\xi}^jD_j(\vartheta_i^r\fett{\xi}^i)
&=&\frac{1}{2}(\partial_i\vartheta_j^r-\partial_j\vartheta_i^r)
\fett{\xi}^i\fett{\xi}^j\\
&=&\frac{1}{2}\left(\vartheta_i^s(\fett{\vartheta}_s\cdot\fett{\partial})
\vartheta_j^r-\vartheta_j^s(\fett{\vartheta}_s\cdot\fett{\partial})
\vartheta_i^r\right)\vartheta_t^i\vartheta_u^j\fett{\vartheta}^t
\fett{\vartheta}^u\\
&=&\frac{1}{2}\left(\vartheta_u^i(\fett{\vartheta}_t\cdot\fett{\partial})
\vartheta_i^r-\vartheta_t^j(\fett{\vartheta}_u\cdot\fett{\partial})
\vartheta_j^r\right)\fett{\vartheta}^t\fett{\vartheta}^u\\
&=&-\frac{1}{2}\left(\vartheta_i^r(\fett{\vartheta}_t\cdot\fett{\partial})
\vartheta_u^i-\vartheta_j^r(\fett{\vartheta}_u\cdot\fett{\partial})
\vartheta_t^j\right)\fett{\vartheta}^t\fett{\vartheta}^u\\
&=&-\frac{1}{2}C_{tu}^r\fett{\vartheta}^t\fett{\vartheta}^u,
\label{dvartheta}
\end{eqnarray}
which is the Maurer-Cartan equation. The exterior derivative of a
general non-coordinate one-form
$\fett{\alpha}=\alpha_r\fett{\vartheta}^r$ is
\begin{equation}
\fett{d}\fett{\alpha}=(\fett{D}\alpha_r)\fett{\vartheta}^r
+\alpha_r\fett{D}\fett{\vartheta}^r
=(\partial_r\alpha_s-\alpha_t\Gamma^t_{rs})
\fett{\vartheta}^r\fett{\vartheta}^s,
\end{equation}
for the exterior derivative of a general $r$-form in the
non-coordinate basis $A^{(r)}=\frac{1}{r!}A_{s_1\ldots
s_r}\fett{\vartheta}^{s_1}\ldots\fett{\vartheta}^{s_r}$ one
obtains
\begin{equation}
\fett{d}A^{(r)}=\frac{(-1)^r}{(r+1)!}
\left(\partial_{\left[s_{r+1}\right.}A_{ \left. s_1\ldots
s_r\right]}-\Gamma^t_{\left[s_{r+1}s_k\right.} A_{\left. s_1\ldots
s_{k-1}ts_{k+1}\ldots s_r\right]}\right)
\fett{\vartheta}^{s_1}\fett{\vartheta}^{s_2}\ldots
\fett{\vartheta}^{s_{r+1}},
\end{equation}
where the square brackets antisymmetrize the lower indices.

The formalism developed so far can also be used to describe tensor
calculus. A tensor is a multilinear map of $r$ vectors and $s$
one-forms into the real numbers and can be written as
\begin{equation}
\mathsf{T}=T^{i_1\ldots i_r}_{j_1\ldots
j_s}\fett{\xi}_{i_1}\otimes
\cdots\otimes\fett{\xi}_{i_r}\otimes\fett{\xi}^{j_1}\otimes\cdots
\otimes\fett{\xi}^{j_s}.
\end{equation}
The components of the tensor are obtained as
\begin{equation}
T^{i_1\ldots i_r}_{j_1\ldots j_s}=\mathsf{T}(\fett{\xi}^{i_1},
\ldots,\fett{\xi}^{i_r},\fett{\xi}_{j_1},\ldots,\fett{\xi}_{j_s})
=\dot{\iota}_{\mbox{\scriptsize
\boldmath$\xi$}^{i_1}\otimes\cdots\otimes\mbox{\scriptsize
\boldmath$\xi$}_{j_s}}\mathsf{T}=(\fett{\xi}^{i_1}\otimes\cdots\otimes
\fett{\xi}_{j_s})\cdot\mathsf{T}.
\end{equation}
For example the metric tensor
$\mathsf{g}=g_{ij}\fett{\xi}^i\otimes\fett{\xi}^j
=g_{ij}\fett{d}x^i\otimes\fett{d}x^j$ maps two vectors
$\fett{a}=a^i\fett{\xi}_i$ and $\fett{b}=b^i \fett{\xi}_i$ into a
scalar according to
\begin{equation}
\mathsf{g}(\fett{a},\fett{b})=\dot{\iota}_{\mbox{\scriptsize
\boldmath$a$}\otimes\mbox{\scriptsize \boldmath$b$}}\mathsf{g}=
(a^k\fett{\xi}_k\otimes b^l \fett{\xi}_l)
\cdot(g_{ij}\fett{\xi}^i\otimes \fett{\xi}^j)=g_{ij}a^ib^j.
\end{equation}

The above tensor concept can be generalized in several ways. For
example one can consider a function that maps $r$ contravariant
and $s$ covariant blades of arbitrary grade into a scalar, i.e.\
tensors of the form
\begin{equation}
\mathsf{T}=T^{i_1\ldots i_r}_{j_1\ldots j_s}A^{(r_1)}_{i_1}\otimes
\cdots\otimes A^{(r_r)}_{i_r}\otimes B_{(s_1)}^{j_1}\otimes\cdots
\otimes B_{(s_s)}^{j_s}.\label{generalizedtensor}
\end{equation}
The other possibility is to consider multivector valued tensors.
In this case a tensor maps a number of (multi)vectors into a
multivector, that does not have to lie in the same vector space.
All these possible generalizations will appear in the following.

\section{Curvature and Torsion}
\setcounter{equation}{0}
Curvature can be described if one transports a vector around a
closed path and measures the difference of the initial and the
transported vector. The path can be thought of as spanned by two
tangent vectors $\fett{a}$ and $\fett{b}$ and closes by
$\left[\fett{a},\fett{b}\right]_{JLB}$. One can then act with a
curvature operator on a tangent vector
$\fett{c}=c^r\fett{\vartheta}_r$:
\begin{multline}
\left[(\fett{a}\cdot\fett{D})(\fett{b}\cdot\fett{D})-
(\fett{b}\cdot\fett{D})(\fett{a}\cdot\fett{D})-\left[\fett{a}
,\fett{b}\right]_{JLB}\cdot\fett{D}\right]\fett{c}\\
=a^rb^sc^t\left(D_rD_s-D_sD_r-C_{rs}^uD_u\right)\fett{\vartheta}_t
=a^rb^sc^tR_{rst}^u\fett{\vartheta}_u,\label{curvop1}
\end{multline}
with
\begin{eqnarray}
R_{rst}^u&=&\left[\big(D_rD_s-D_sD_r-\left[\fett{\vartheta}_r,
\fett{\vartheta}_s\right]_{JLB}\cdot\fett{D}\big)\fett{\vartheta}_t
\right]\cdot\fett{\vartheta}^u\\
&=&\left[D_r(\Gamma_{st}^w\fett{\vartheta}_w)-D_s(\Gamma_{rt}^w
\fett{\vartheta}_w)-C_{rs}^w(D_w\fett{\vartheta}_t)\right]\cdot
\fett{\vartheta}^u\\
&=&\partial_r\Gamma_{st}^u-\partial_s\Gamma_{rt}^u
+\Gamma_{rw}^u\Gamma_{st}^w-\Gamma_{sw}^u\Gamma_{rt}^w
+\Gamma_{rs}^w\Gamma_{wt}^u-\Gamma_{sr}^w\Gamma_{wt}^u,\label{Rurstdef}
\end{eqnarray}
which in the case of a coordinate basis reduces to
\begin{equation}
R_{ijk}^l=\left[\big(D_iD_j-D_jD_i\big)\fett{\xi}_k
\right]\cdot\fett{\xi}^l
=\partial_i\Gamma_{jk}^l-\partial_j\Gamma_{ik}^l
+\Gamma_{im}^l\Gamma_{jk}^m-\Gamma_{jm}^l\Gamma_{ik}^m.
\end{equation}
Since the curvature operator maps three vectors into a fourth one,
it can also be written as a tensor
$\mathsf{R}=R_{rst}^u\fett{\vartheta}_u\otimes\fett{\vartheta}^r
\otimes\fett{\vartheta}^s\otimes\fett{\vartheta}^t$. In general
the curvature operator can act on a multivector $A$, so that one
has with (\ref{partimultvec})
\begin{multline}
\big[(\fett{a}\cdot\fett{D})(\fett{b}\cdot\fett{D})-
(\fett{b}\cdot\fett{D})(\fett{a}\cdot\fett{D})-\big[\fett{a}
,\fett{b}\big]_{JLB}\cdot\fett{D}\big]A\\
=\big[(\fett{a}\cdot\fett{\partial})\mathtt{S}(\fett{b})
-(\fett{b}\cdot\fett{\partial})\mathtt{S}(\fett{a})
+\mathtt{S}(\fett{a})\times \mathtt{S}(\fett{b})
-\mathtt{S}(\left[\fett{a},\fett{b}\right]_{JLB})\big]\times A
=\mathtt{R}(\fett{ab})\times A,\label{Rabdef}
\end{multline}
which reduces to
\begin{equation}
\left[(\fett{a}\cdot\fett{D})(\fett{b}\cdot\fett{D})-
(\fett{b}\cdot\fett{D})(\fett{a}\cdot\fett{D})-\left[\fett{a}
,\fett{b}\right]_{JLB}\cdot\fett{D}\right]\fett{c}
=\mathtt{R}(\fett{ab})\cdot\fett{c}\label{Rttdef}
\end{equation}
acting on a vector. The bivector-valued function of a bivector
\begin{equation}
\mathtt{R}(\fett{ab})=(\fett{a}\cdot\fett{\partial})\mathtt{S}(\fett{b})
-(\fett{b}\cdot\fett{\partial})\mathtt{S}(\fett{a})
+\mathtt{S}(\fett{a})\times \mathtt{S}(\fett{b})
-\mathtt{S}(\left[\fett{a},\fett{b}\right]_{JLB})
\end{equation}
fulfills the Ricci and Bianchi identities
\begin{eqnarray}
&&\fett{a}\cdot\mathtt{R}(\fett{bc})+\fett{b}\cdot\mathtt{R}(\fett{ca})
+\fett{c}\cdot\mathtt{R}(\fett{ab})=0\\
\mathrm{and}\quad&&(\fett{a}\cdot\fett{D})\mathtt{R}(\fett{bc})
+(\fett{b}\cdot\fett{D})\mathtt{R}(\fett{ca})
+(\fett{c}\cdot\fett{D})\mathtt{R}(\fett{ab})=0.
\end{eqnarray}

Comparing (\ref{curvop1}) with (\ref{Rttdef}) shows that the
curvature is described by a bivector-valued function of a
bivector according to
\begin{equation}
a^rb^sc^tR_{rst}^u\fett{\vartheta}_u=\mathtt{R}(\fett{ab})\cdot\fett{c}.
\end{equation}
But it is also possible to describe it by a scalar-valued function
of a bivector, i.e.\ a two-form $R_t^u(\fett{ab})
=\dot{\iota}_{\mbox{\scriptsize\boldmath$ab$}}R_t^u$ according to
\begin{equation}
a^rb^sc^tR_{rst}^u\fett{\vartheta}_u=c^tR_t^u(\fett{ab})\fett{\vartheta}_u.
\label{Rtudef}
\end{equation}
It is now easy to see from this definition and (\ref{Rurstdef})
that the curvature two-form $R_t^u$ has the form
\begin{equation}
R_t^u=\big(\partial_v\Gamma_{wt}^u +\Gamma_{rt}^u\Gamma^r_{wv}
+\Gamma^u_{vr}\Gamma^r_{wt}\big)\fett{\vartheta}^v\fett{\vartheta}^w,
\label{Rtucompo}
\end{equation}
which also can be expressed in another way. For this purpose one notices
that the exterior derivative of $\fett{\vartheta}_r$ is a
vector-valued one-form:
\begin{equation}
\fett{d}\fett{\vartheta}_r=\fett{\vartheta}^sD_s\fett{\vartheta}_r
=\Gamma_{sr}^t\fett{\vartheta}^s\fett{\vartheta}_t
=\fett{\omega}_r^t\fett{\vartheta}_t,\label{dvarthetar}
\end{equation}
where $\fett{\omega}_r^t=\Gamma_{sr}^t\fett{\vartheta}^s$. With
$\fett{\omega}_r^t$ the curvature two-form (\ref{Rtucompo}) can
also be written as
\begin{equation}
R_t^u=\fett{d}\fett{\omega}_t^u+\fett{\omega}_r^u\fett{\omega}^r_t,
\label{Cartanstrucfirst}
\end{equation}
which is the first Cartan structure equation. Exterior
differentiation of (\ref{Cartanstrucfirst}) gives the Bianchi
identity for the curvature two-form:
$\fett{d}R_s^r+\fett{\omega}_t^rR_s^t-R_t^r\fett{\omega}_s^t=0$.

It is possible that the path spanned by two tangent vectors
$\fett{a}$ and $\fett{b}$ is not closed by
$\left[\fett{a},\fett{b}\right]_{JLB}$. This is measured by the
torsion
\begin{equation}
(\fett{a}\cdot\fett{D})\fett{b}-(\fett{b}\cdot\fett{D})\fett{a}-
\left[\fett{a},\fett{b}\right]_{JLB}=a^rb^sT^t_{rs}\fett{\vartheta}_t
\end{equation}
with
\begin{equation}
T^t_{rs}=\big[D_r\fett{\vartheta}_s-D_s\fett{\vartheta}_r-
\left[\fett{\vartheta}_r,\fett{\vartheta}_s\right]_{JLB}\big]\cdot\fett{\vartheta}^t
=\Gamma_{rs}^t-\Gamma_{sr}^t-C_{rs}^t,\label{Ttrsdef}
\end{equation}
which reduces in a coordinate basis to $T_{ij}^k
=\Gamma_{ij}^k-\Gamma_{ji}^k$. This means that for non-vanishing
torsion the $\Gamma_{ij}^k$ are no longer symmetric in the lower
indices so that $\fett{dd}x^i$ is no longer zero and the exterior
differential of an $r$-form is given by
\begin{eqnarray}
\fett{d}A^{(r)}=\fett{D}A^{(r)}
&=&\frac{1}{r!}\left(\frac{\partial A_{i_1i_2\ldots i_r}}
{\partial x^j}\right)\fett{D}x^j\fett{D}x^{i_1}\fett{D}x^{i_2}
\ldots\fett{D}x^{i_r}\nonumber\\
&&+\frac{1}{r!}A_{i_1i_2\ldots i_r}
\left[\fett{DD}x^{i_1}\fett{D}x^{i_2}\ldots\fett{D}x^{i_r}
-\fett{D}x^{i_1}\fett{DD}x^{i_2}\fett{D}x^{i_3}\ldots\fett{D}x^{i_r}\right.
\nonumber\\
&&\hphantom{+\frac{1}{r!}A_{i_1i_2\ldots i_r}\;}\left.
+\ldots+(-1)^{r-1}\fett{D}x^{i_1}\fett{D}x^{i_2}\ldots\fett{DD}x^{i_r}
\right].
\end{eqnarray}

The torsion maps two vectors into a third one, so that it can also
be written as a tensor
$\mathsf{T}=T^t_{rs}\fett{\vartheta}_t\otimes\fett{\vartheta}^r
\otimes\fett{\vartheta}^s$. The other possibility is to describe
the torsion with a scalar-valued function of a bivector, i.e.\ a
two-form $T^t(\fett{ab})= \dot{\iota}_{{\mbox{\scriptsize
\boldmath$ab$}}}T^t$ according to
\begin{equation}
a^rb^sT^t_{rs}\fett{\vartheta}_t=T^t(\fett{ab})\fett{\vartheta}_t.
\end{equation}
It is then easy to see with (\ref{Ttrsdef}) that the torsion
two-form can be written as
\begin{equation}
T^t=\left(\Gamma_{rs}^t-\frac{1}{2}C^t_{sr}\right)
\fett{\vartheta}^r\fett{\vartheta}^s.
\end{equation}
With the Cartan one-form $\fett{\omega}_r^t$ this can also be
written as
\begin{equation}
T^t=\fett{d}\fett{\vartheta}^t+\fett{\omega}_r^t\fett{\vartheta}^r,
\label{Cartanstruczwei}
\end{equation}
which is the second Cartan structure equation. Applying the
exterior differentiation on both sides of (\ref{Cartanstruczwei})
gives the second Bianchi-identity
$\fett{d}T^t+\fett{\omega}_r^tT^r=R_r^t\fett{\vartheta}^r$.

\section{Rotor Groups and Bivector Algebras}
\setcounter{equation}{0}
The multivectors of even Grassmann grade are closed under the
Clifford star product and form the group $Spin(p,q)$. An element
$S\in Spin(p,q)$ fulfills $S\starP\overline{S}=\pm 1$ and a
transformation $S\starP\fett{x}\starP S^{-1\starP}$ gives again a
vector-valued result \cite{Doran1}. The elements $R\in Spin(p,q)$
with $R\starP\overline{R}=+1$ are called rotors and form the rotor
group $Spin^+(p,q)$, which in the euclidian case is equal to the
spin-group. For a rotor one has $R^{-1\starP}=\overline{R}$, so
that a multivector $A$ transforms as $R\starP
A\starP\overline{R}$. A rotor can be written as a starexponential
of a bivector, i.e.\ in general the rotor has for a bivector
$\mathtt{B}$ the form
\begin{equation}
R(t)=\pm e_{\starP}^{\frac{t}{2}\mathtt{B}}
\end{equation}
and the rotation of a vector $\fett{x}_0$ is given by $\fett{x}(t)
=R(t)\starP\fett{x}_0\starP\overline{R}(t)$. The bivector basis
$\mathtt{B}_i$ of a rotor constitutes an algebra under the
commutator product
\begin{equation}
\mathtt{B}_i\times\mathtt{B}_j=C_{ij}^k\mathtt{B}_k,
\label{bivecalgebra}
\end{equation}
where the $C_{ij}^k$ are the structure constants (note that one
has here an additional factor $\frac{1}{2}$ due to the definition
of the commutator product). Furthermore one can directly calculate
\begin{equation}
\kappa_{ij}=\mathtt{B}_i\cdot \mathtt{B}_j,
\end{equation}
which is (proportional to) the Killing metric. As an
example one can consider the group $SO(3)$. Given a three
dimensional euclidian space with basis vectors $\fett{\sigma}_i$
the rotor is given by
\begin{equation}
R=R_0+R_1\fett{\sigma}_2\fett{\sigma}_3+R_2\fett{\sigma}_3\fett{\sigma}_1
+R_3\fett{\sigma}_1\fett{\sigma}_2,\label{rotorRi}
\end{equation}
with $R\starP\overline{R}=R_0^2+R_1^2+R_2^2+R_3^2=1$, so that the
rotor can also be parametrized with three parameters $\alpha$,
$\theta$ and $\varphi$ as:
\begin{equation}
R(\alpha,\theta,\varphi)=\cos\alpha\cos\theta+\sin\alpha\cos\varphi\fett{\sigma}_2\fett{\sigma}_3
+\sin\alpha\sin\varphi\fett{\sigma}_3\fett{\sigma}_1+\cos\alpha
\sin\theta\fett{\sigma}_1\fett{\sigma}_2.\label{SO3R}
\end{equation}
The three basis bivectors
$\mathtt{B}_1=\fett{\sigma}_2\fett{\sigma}_3$,
$\mathtt{B}_2=\fett{\sigma}_3\fett{\sigma}_1$ and
$\mathtt{B}_3=\fett{\sigma}_1\fett{\sigma}_2$ fulfill
\begin{equation}
\mathtt{B}_i\times \mathtt{B}_j=-\varepsilon_{ijk}\mathtt{B}_k
\qquad\mathrm{and}\qquad \kappa_{ij}=\mathtt{B}_i\cdot
\mathtt{B}_j=-\delta_{ij}. \label{SO3bivecalg}
\end{equation}
It is easy to see that the group vector manifold, which for
$SO(3)$ is an $S^3$ embedded in a four dimensional euclidian space
with basis vectors $\fett{\tau}_a$, can be read off from
(\ref{SO3R}) as
\begin{equation}
\fett{r}_R(\alpha,\theta,\varphi)=\cos\alpha\cos\theta\,\fett{\tau}_1
+\sin\alpha\cos\varphi\,\fett{\tau}_2
+\sin\alpha\sin\varphi\,\fett{\tau}_3+\cos\alpha
\sin\theta\,\fett{\tau}_4.
\end{equation}

The rotors act on themselves by left- and right-translation. A
left-translation with a rotor $R'$ is given by
$\ell_{R'}R=R'\starP R$ and on the group vector manifold by
$\ell_{R'}\fett{r}_{R}=\fett{r}_{R'\starP R}$. The
left-translation induces a map $T_R\ell_{R'}$ between the tangent
spaces at $\fett{r}_R$ and $\fett{r}_{R'\starP R}$. A vector field
$\fett{a}(\fett{r}_R)$ on the group vector manifold is
left invariant if $T_{R}\ell_{R'}\fett{a}(\fett{r}_{R})
=\fett{a}(\fett{r}_{R'\starP R})$. Left invariant vector fields on
the group vector manifold can be obtained if one considers the
multivector fields on the rotors given by
$B^{\mathrm{left}}_i(R)=R\starP\mathtt{B}_i$. For two rotors $R$
and $R'$ one has
\begin{equation}
B^{\mathrm{left}}_i(R'\starP R)=R'\starP B^{\mathrm{left}}_i(R).
\label{Bleft}
\end{equation}
Just as to each rotor $R$ in the $\fett{\sigma}_a$-space
corresponds a vector $\fett{r}_R$ in the $\fett{\tau}_a$-space
there is also for each multivector field $B^{\mathrm{left}}_i(R)$
in the $\fett{\sigma}_a$-space a left invariant vector field
$\fett{\vartheta}_{B^{\mathrm{left}}_i(R)}(\fett{r}_R)\equiv
\fett{\vartheta}_i$ in the $\fett{\tau}_a$-space. These vector
fields are closed under the Jacobi-Lie-bracket, i.e.\ they form a
Lie subalgebra of all vector fields on $\fett{r}_R$ and they form
a non-coordinate basis on $\fett{r}_R$, for the $SO(3)$-case one
has for example $\fett{\vartheta}_i\cdot\fett{\vartheta}_j
=\delta_{ij}$.
The multivector fields $B^{\mathrm{left}}_i(R)$ are uniquely
defined by the bivectors at $R=1$ and the corresponding
left invariant vector fields are uniquely defined by their value
in $\fett{r}_{R=1}$. In the $SO(3)$-example the tangent space at
$\fett{r}_{R=1}(0,0,0)=\fett{\tau}_1$ is spanned by the vectors
$\fett{\vartheta}_{\mathtt{B}_i}=\fett{\tau}_{i+1}$ and
constitutes the $\mathfrak{so}(3)$ algebra in the
$\fett{\tau}_a$-space, where the commutator product in the
bivector algebra corresponds here in the $\mathfrak{so}(3)$-case
to the vector cross product on the
$\fett{\vartheta}_{\mathtt{B}_i}$-space, i.e.
\begin{equation}
\fett{\vartheta}_{\mathtt{B}_i\times\mathtt{B}_j}
=-\fett{\vartheta}_{\mathtt{B}_i}\times\fett{\vartheta}_{\mathtt{B}_j}.
\end{equation}

To each basis-bivector $\mathtt{B}_i$ of the bivector algebra a
two-form $\Theta^i$ can be found so that
$\dot{\iota}_{\mathtt{B}_i}\Theta^j=\overline{\mathtt{B}_i}\cdot\Theta^j
=\delta_i^j$ and to the two-forms $\Theta^i$ correspond then in
the $\fett{\tau}_a$-space one-forms $\fett{\vartheta}^{\Theta^i}
\equiv\fett{\vartheta}^i$ that generalize to reciprocal
non-coordinate basis vector fields on $\fett{r}_R$, which clearly
obey the Maurer-Cartan equation (\ref{dvartheta}). For a $r$-form
$\fett{A}^{(r)}$ on the group vector manifold that is
vector-valued in the $\fett{\sigma}_a$-space one can then in
analogy to (\ref{iAr}) and with $\dot{\iota}_{\mbox{\scriptsize
\boldmath$\vartheta$}_1\ldots\mbox{\scriptsize
\boldmath$\vartheta$}_r}\fett{A}^{(r)}=\fett{A}^{(r)}
(\fett{\vartheta}_1\ldots\fett{\vartheta}_r)$ define the
BRST-operator $s$ as
\begin{eqnarray}
\big(s\fett{A}^{(r)}\big)\left(\fett{\vartheta}_{1}\fett{\vartheta}_{2}
\ldots\fett{\vartheta}_{r+1}\right)
&=&\sum_{n=1}^{r+1}(-1)^{n+1}\mathtt{B}_n\cdot\fett{A}^{(r)}\left(
\fett{\vartheta}_1\ldots\check{\fett{\vartheta}}_n
\ldots\fett{\vartheta}_{r+1}\right)\nonumber\\
&&+\sum_{m<n}(-1)^{m+n}\fett{A}^{(r)}\left(\left[\fett{\vartheta}_m,
\fett{\vartheta}_n\right]_{JLB}
\fett{\vartheta}_1\ldots\check{\fett{\vartheta}}_m\ldots
\check{\fett{\vartheta}}_n \ldots\fett{\vartheta}_{r+1}\right).
\end{eqnarray}
The $s$-operator can then be written as (see for example
\cite{Azcarraga} and the references therein):
\begin{equation}
s=\mathtt{B}_i\cdot\otimes\fett{\vartheta}^i+\frac{1}{2}C_{ij}^k\fett{\vartheta}^j
\fett{\vartheta}^i\frac{\partial}{\partial\fett{\vartheta}^k}.
\end{equation}

The adjoint action of the rotor group on the bivector algebra is
given by \cite{Doran1}
\begin{equation}
\mathrm{Ad}_R \mathtt{B}=R\starP
\mathtt{B}\starP\overline{R},\label{AdRB}
\end{equation}
where $\mathtt{B}=b^i\mathtt{B}_i$ is a general element of the
bivector algebra, to which in the
$\fett{\vartheta}_{\mathtt{B}_i}$-space corresponds a vector
$\fett{b}=b^i\fett{\vartheta}_{\mathtt{B}_i}$. $\mathrm{Ad}_R$ is
a bivector algebra homomorphism, i.e.
$\mathrm{Ad}_R(\mathtt{A}\times\mathtt{B})=\mathrm{Ad}_R\mathtt{A}
\times\mathrm{Ad}_R\mathtt{B}$ and a left action, i.e.
$\mathrm{Ad}_{R\starP R'}=\mathrm{Ad}_R\mathrm{Ad}_{R'}$. For all
elements $R$ of the rotor group the adjoint action (\ref{AdRB})
constitutes the adjoint bivector orbit of $\mathtt{B}$, to which
in the $\fett{\vartheta}_{\mathtt{B}_i}$-space corresponds an
orbit vector manifold. In the $SO(3)$-case the adjoint action
(\ref{AdRB}) leaves $|\mathtt{B}|^2=\sum_{i=1}^3(b^i)^2
=|\fett{b}|^2$ invariant, so that the adjoint orbit vector
manifold is an $S^2$.

Let now $\mathtt{A}$ be an element of the bivector algebra and
consider the rotor $R(t)=e_{\starP}^{\frac{t}{2}\mathtt{A}}$. The
adjoint action of this one-parameter rotor subgroup gives a curve
in the bivector orbit, and the derivative at $t=0$ is
\begin{equation}
\mathrm{ad}_{\mathtt{A}}\mathtt{B}=\left.\frac{d}{dt}\right|_{t=0}
R(t)\starP\mathtt{B}\starP\overline{R(t)}=\mathtt{A}\times
\mathtt{B}.
\end{equation}
In the $\fett{\vartheta}_{\mathtt{B}_i}$-space the vector
$\fett{\vartheta}_{\mathtt{A}\times\mathtt{B}}$ is the tangent
vector in direction $\fett{\vartheta}_{\mathtt{A}}$ to the orbit
vector manifold in the point $\fett{\vartheta}_{\mathtt{B}}$,
i.e.\ $\fett{\vartheta}_{\mathtt{A}\times\mathtt{B}}$ generates
the adjoint action corresponding to $\mathtt{A}$. It is also
possible to define the coadjoint action $\mathrm{Ad}^*_R$ of the
rotor group on a two-form $\Theta$ by
\begin{equation}
\overline{\mathtt{B}}\cdot\mathrm{Ad}^*_R\Theta
=\overline{\mathrm{Ad}_{R}\mathtt{B}}\cdot\Theta, \label{coAdDef}
\end{equation}
which is the right action $\mathrm{Ad}^*_R\Theta
=\overline{R}\starP\Theta\starP R$. The coadjoint left action is
given by $\mathrm{Ad}^*_{\overline{R}}\Theta$. Infinitesimally one
has $\overline{\mathtt{B}}\cdot\mathrm{ad}_{\mathtt{A}}^*\Theta
=\overline{\mathrm{ad}_{\mathtt{A}}\mathtt{B}}\cdot\Theta$, or
$\mathrm{ad}^*_{\mathtt{A}}\Theta=\Theta\times\mathtt{A}$. In the
$SO(3)$-case the rotor acts on an euclidian space where the basis
vectors and the reciprocal basis vectors are actually the same, so
that $\mathtt{B}_i=\Theta^i$ and there is no difference between
the adjoint and the coadjoint action.

In the above discussion the rotor $R$ acts intrinsically from the
left on a vector space. But more generally a rotor in an ambient
space can also act from the left on a vector manifold
$\fett{x}(x^i)$ by $\fett{x}'=R\starP \fett{x}\starP\overline{R}$
if $\fett{x}'$ is again a point in the vector manifold. The
left-action of the rotor $R(t)=e_{\starP}^{\frac{t}{2}\mathtt{B}}$
induces on the vector manifold $\fett{x}(x^i)$ the vector field
\begin{equation}
\left.\frac{d}{dt}\right|_{t=0}R(t)\starP\fett{x}\starP
\overline{R(t)}=\mathtt{B}\cdot\fett{x}. \label{RxRt0}
\end{equation}
Furthermore a short calculation shows that there is an algebra
anti-homomorphism between the bivector algebra in the ambient
space and the induced vector fields on the vector manifold, given
by
\begin{equation}
\left[\mathtt{A}\cdot\fett{x},\mathtt{B}\cdot\fett{x}\right]_{JLB}
=-(\mathtt{A}\times\mathtt{B})\cdot\fett{x}.\label{xAxBJLBxAB}
\end{equation}

The rotor in the ambient space acts not only on the vectors
$\fett{x}$ of the vector manifold, but in the same way also on
tangent vectors $\fett{a}$ at the manifold which are vectors in
the ambient space too. The transformation of $\fett{x}$ and
$\fett{a}$ in the ambient space of the vector manifold induce a
transformation in the tangent bundle. The tangent bundle manifold
can be seen as a $2d$-dimensional vector manifold in a
$(2d+2)$-dimensional ambient space with basis vectors
$\fett{\sigma}_a$ and $\fett{\tau}_a$, i.e.\ as
\begin{equation}
(\fett{x}+\fett{a})(x^i,a^i) =x^a(x^i)\fett{\sigma}_a
+a^j\xi_j^a(x^i)\fett{\tau}_a.
\end{equation}
Analogously one can define multivector bundles, for example a
bivector bundle manifold has the form
\begin{equation}
(\fett{x}+\mathtt{B})(x^i,B^{jk})=x^a(x^i)\fett{\sigma}_a+B^{jk}
\xi^a_j(x^i)\xi^b_k(x^i)\fett{\tau}_a\fett{\tau}_b.
\end{equation}
The tangential lift of the rotor action is given by
$R\starP\fett{x}\starP\overline{R}
+R\starP\fett{a}\starP\overline{R}$, where the rotor acts on the
$\fett{\tau}_a$-space in the same way as on the
$\fett{\sigma}_a$-space. In the case of a flat vector manifold the
tangent bundle is just a $2d$-dimensional vector space and the
rotor acts separately and intrinsically on both subspaces. Instead
of two rotors that act separately on the $\fett{\sigma}_a$ and
$\fett{\tau}_a$ spaces one can consider also a lifted rotor with a
bivector $\mathtt{B}_{\mathrm{lifted}}$ that is the sum of the two
single bivectors, so that one can write $R_{\mathrm{lifted}}\starP
(\fett{x}+\fett{a})\starP\overline{R_{\mathrm{lifted}}}$. If one
describes the tangent vector in a reciprocal ambient space, i.e.\
as a one-form $\fett{\alpha}$ the cotangent bundle has the form
\begin{equation}
(\fett{x}+\fett{\alpha})(x^i,\alpha_i) =x^a(x^i)\fett{\sigma}_a
+\alpha_i\xi_a^i(x^i)\fett{\tau}^a
\end{equation}
and the corresponding cotangent lift is given by
$\overline{R}\starP\fett{x}\starP R
+\overline{R}\starP\fett{\alpha}\starP R$ or
$\overline{R_{\mathrm{lifted}}}\starP(\fett{x}+\fett{\alpha})\starP
R_{\mathrm{lifted}}$.

In order to construct unitary transformations \cite{Doran3} one
considers a $2n$-dimensional space with basis vectors
$\fett{\alpha}_i$ and $\fett{\beta}_i$ for $i=1,\ldots,n$ and a
bivector
\begin{equation}
\mathtt{J}=\sum_{i=1}^d\fett{\alpha}_i\fett{\beta}_i=\sum_{i=1}^d
\mathtt{J}_i.
\end{equation}
The two subspaces spanned by $\fett{\alpha}_i$ and
$\fett{\beta}_i$ should have the same metric, i.e.\
$\fett{\alpha}_i\cdot\fett{\alpha}_j
=\fett{\beta}_i\cdot\fett{\beta}_j$ and $\fett{\alpha}_i\cdot
\fett{\beta}_j=0$. The $2n$-dimensional vector
$\fett{x}=a^i\fett{\alpha}_i +b^i\fett{\beta}_i$ corresponds to an
$n$-dimensional complex vector with components
$x^k=\fett{x}\cdot\fett{\alpha}_k+\iu\,\fett{x}\cdot\fett{\beta}_k
=a^k+\iu\, b^k$ and the complex internal product can be written as
\begin{equation}
\langle\fett{x}|\fett{y}\rangle=x^k\overline{y}_k
=(\fett{x}\cdot\fett{\alpha}^k+\iu\,\fett{x}\cdot\fett{\beta}^k)
(\fett{y}\cdot\fett{\alpha}_k-\iu\,\fett{y}\cdot\fett{\beta}_k)
=\fett{x}\cdot\fett{y}+\iu(\fett{xy})\cdot\mathtt{J}.
\label{complprod}
\end{equation}
A unitary transformation generated by the rotor $R$ leaves the
above complex product invariant, i.e.
\begin{equation}
(\fett{xy})\cdot\mathtt{J}=\big((R\starP\fett{x}\starP\overline{R})
(R\starP\fett{y}\starP\overline{R})\big)\cdot\mathtt{J}
=(\fett{xy})\cdot(\overline{R}\starP\mathtt{J}\starP R),
\end{equation}
which means that $\mathtt{J}=R\starP\mathtt{J}\starP\overline{R}$
is the defining relation for the unitary rotor. With the ansatz
$R=e_{\starP}^{\mathtt{B}/2}$ one obtains the defining relation
for the bivector $\mathtt{B}$
\begin{equation}
\mathtt{B}\times\mathtt{J}=0,
\end{equation}
which is solved by $\mathtt{B}
=\fett{xy}+(\fett{x}\cdot\mathtt{J})(\fett{y}\cdot \mathtt{J})$
Putting in this formula the basis vectors for $\fett{x}$ and
$\fett{y}$ one obtains the $n^2$ basis bivectors of the
$\mathfrak{u}(n)$-algebra:
\begin{equation}
\mathtt{E}_{ij}=\fett{\alpha}_i\fett{\alpha}_j+\fett{\beta}_i\fett{\beta}_j,
\qquad \mathtt{F}_{ij}=\fett{\alpha}_i\fett{\beta}_j
-\fett{\beta}_i\fett{\alpha}_j \qquad\mathrm{and}\qquad
\mathtt{J}_i=\fett{\alpha}_i\fett{\beta}_i
\end{equation}
for $i<j=1,\ldots,d$. It is easy to show that these basis
bivectors form a closed algebra under the commutator product. The
bivector $\mathtt{J}$ is part of the $\mathfrak{u}(n)$-algebra, if
one excludes this generator of a global phase one obtains the
$\mathfrak{su}(n)$-algebra.

In order to describe the $Gl(n)$ by rotors one proceeds in a way similar to
the unitary case. One considers a $2n$-dimensional space spanned
by the basis vectors $\fett{\alpha}_i$ and $\fett{\beta}_i$ for
$i=1,\ldots,n$, but now the metric in the spaces spanned by
$\fett{\alpha}_i$ and $\fett{\beta}_i$ is opposite, i.e.\ the
Clifford star product is given by
\begin{equation}
\starP=\exp\left[\eta_{ij}
\frac{\overleftarrow{\partial}}{\partial\fett{\alpha}_i}
\frac{\overrightarrow{\partial}}{\partial\fett{\alpha}_j}-\eta_{ij}
\frac{\overleftarrow{\partial}}{\partial\fett{\beta}_i}
\frac{\overrightarrow{\partial}}{\partial\fett{\beta}_j}\right].\label{glnstar}
\end{equation}
On this space a bivector
$\mathtt{K}=\fett{\alpha}_i\fett{\beta}^i$ can be defined, so that
one can decompose a vector $\fett{x}$ according to
\begin{equation}
\fett{x}=\frac{1}{2}\big(\fett{x}+\fett{x}\cdot\mathtt{K}\big)+
\frac{1}{2}\big(\fett{x}-\fett{x}\cdot\mathtt{K}\big)
=\fett{x}_++\fett{x}_-,
\end{equation}
so that $\fett{x}_+\cdot\fett{x}_+=\fett{x}_-\cdot\fett{x}_-=0$.
There are then two subspaces $V_+$ and $V_-$ defined by
$\fett{x}_+\cdot\mathtt{K}=\fett{x}_+$ and
$\fett{x}_-\cdot\mathtt{K}=\fett{x}_-$. A $Gl(n)$-transformation
transforms now a vector in $V_+$ into another vector in $V_+$,
i.e.
\begin{equation}
(R\starP \fett{x}_+\starP \overline{R})\cdot\mathtt{K}
=R\starP\fett{x}_+ \starP\overline{R},
\end{equation}
or $\mathtt{K}=R\starP\mathtt{K}\starP\overline{R}$. With the same
argumentation as above one can see that a bivector generator must
have the form
$\mathtt{B}=\fett{x}\fett{y}-(\fett{x}\cdot\mathtt{K})
(\fett{y}\cdot\mathtt{K})$, so that the $n^2$ basis bivectors of
$\mathfrak{gl}(n)$ are
\begin{equation}
\mathtt{E}_{ij}=\fett{\alpha}_i\fett{\alpha}_j-\fett{\beta}_i\fett{\beta}_j,
\qquad \mathtt{F}_{ij}=\fett{\alpha}_i\fett{\beta}_j
-\fett{\beta}_i\fett{\alpha}_j\qquad\mathrm{and}\qquad
\mathtt{K}_i=\fett{\alpha}_i\fett{\beta}_i \label{glngenerators}
\end{equation}
for $i<j=1,\ldots,n$.

The $Gl(n)$-case can be understood in another way if one transforms
the variables of the vector
$\fett{x}=a^i\fett{\alpha}_i+b^i\fett{\beta}_i$ into variables
$q^i,\, p^i,\, \fett{\eta}_i$ and $\fett{\rho}_i$ according to
\begin{eqnarray}
\fett{x}_+&=&\frac{1}{2}\big(\fett{x}+\fett{x}\cdot\mathtt{K}\big)
=\frac{1}{2}(a^i-b^i)(\fett{\alpha}_i-\fett{\beta}_i)
\equiv q^i\fett{\eta}_i\\
\fett{x}_-&=&\frac{1}{2}\big(\fett{x}-\fett{x}\cdot\mathtt{K}\big)
=\frac{1}{2}(a^i+b^i)(\fett{\alpha}_i+\fett{\beta}_i) \equiv
p^i\fett{\rho}_i.
\end{eqnarray}
It is then straight forward to transform the star product
(\ref{glnstar}) and the generators (\ref{glngenerators}) into
these new variables. For the star product one obtains
\begin{equation}
\starP=\exp\left[\frac{\eta_{ij}}{2}\left(
\frac{\overleftarrow{\partial}}{\partial\fett{\eta}_i}
\frac{\overrightarrow{\partial}}{\partial\fett{\rho}_j}+
\frac{\overleftarrow{\partial}}{\partial\fett{\rho}_i}
\frac{\overrightarrow{\partial}}{\partial\fett{\eta}_j}\right)\right],
\end{equation}
which is a fermionic version of the Moyal product
\begin{equation}
\starM=\exp\left[\frac{\iu\hbar}{2}\eta^{ij}
\left(\frac{\overleftarrow{\partial}}{\partial q^i}
\frac{\overrightarrow{\partial}}{\partial p^j}
-\frac{\overleftarrow{\partial}}{\partial p^i}
\frac{\overrightarrow{\partial}}{\partial q^j}\right)\right].
\end{equation}
This suggests that the vector
$\fett{x}=q^i\fett{\eta}_i+p^i\fett{\rho}_i$ can not only be
transformed with a fermionic star exponential as described above,
but can also be transformed in the bosonic coefficients with a
bosonic star exponential according to \cite{Arnal}
\begin{equation}
e_{\starM}^{\alpha_{ij}M^{ij}}\starM q^k \starM
e_{\starM}^{-\alpha_{ij}M^{ij}}=q^k+\alpha_{ij}\left[M^{ij},q^k\right]_{\starM}
+\frac{1}{2!}\alpha_{ij}\alpha_{lm}\big[M^{lm},\left[M^{ij},q^k\right]_{\starM}
\big]_{\starM}+\ldots ,
\end{equation}
where $\left[f,g\right]_{\starM}=f\starM g-g\starM f$ is the
star-commutator. In analogy to the fermionic case one can now
demand that for a $Gl(n)$ transformation the $q^k$ have to be a
linear combination of the $q^i$ alone and no terms in $p^i$ should
appear. This means that $\left[M^{ij},q^k\right]_{\starM}$ must be
a function of the $q^i$ alone. This is achieved if one chooses the
bosonic generators
\begin{equation}
E^{ij}=q^ip^j+q^jp^i,\qquad
F^{ij}=q^ip^j-q^jp^i,\qquad\mathrm{and}\qquad K^i=q^ip^i,
\end{equation}
which form a closed algebra under the Moyal star-commutator.

\section{Active and passive Rotations and the theoretical Prediction of Spin}
\setcounter{equation}{0}
A general multivector is now invariant under a combined
transformation of the bosonic coefficients and a compensating
transformation of the fermionic basis vectors. The bosonic
transformation of the coefficients is an active transformation and
the fermionic transformation of the basis vectors is a passive
transformation. In a tuple formalism this difference cannot be
made and so active and passive transformations are mixed up with
left and right transformations, whereas in a multivector formalism
one rather has that an active right transformation corresponds to
a passive left transformation, and the other way round.

To illustrate the concept of active and passive transformations in
the star product formalism one can consider rotations in space and
space-time. In the three dimensional euclidian space with vectors
$\fett{x}=x^i\fett{\sigma}_i$ the active rotations \cite{Arnal}
are generated by the angular momentum functions
\begin{equation}
L^i=\varepsilon^{ijk}x^jp^k,
\end{equation}
which fulfill with the three dimensional Moyal product
\begin{equation}
\starM=\exp\left[\frac{\iu\hbar}{2}\sum_{i=1}^3
\left(\frac{\overleftarrow{\partial}}{\partial x^{i}}
\frac{\overrightarrow{\partial}}{\partial p^{i}}
-\frac{\overleftarrow{\partial}}{\partial p^{i}}
\frac{\overrightarrow{\partial}}{\partial x^{i}}\right)\right]
\end{equation}
the active algebra
\begin{equation}
\left[L^i,L^j\right]_{\starM}=\iu\hbar\varepsilon^{ijk}L^k.
\end{equation}
An active left-rotation has then the form
\begin{equation}
\fett{x}'=\overline{U}\starM\fett{x}\starM
U=e_{\starM}^{-\frac{\iu}{\hbar}\alpha_{k}L^{k}} \starM
\fett{x}\starM e_{\starM}^{\frac{\iu}{\hbar}\alpha_{k}L^{k}}
=\big(R^{i}_{j}x^{j}\big)\fett{\sigma}_{i},
\end{equation}
where the $R^i_j$ is the well known rotation matrix. The
corresponding passive rotation \cite{Doran3,Ashdown} is generated
by the bivectors
\begin{equation}
\mathtt{B}_i=\frac{1}{2}\varepsilon_{ijk}\fett{\sigma}_j\fett{\sigma}_k
\end{equation}
that fulfill as seen above the passive algebra
\begin{equation}
\mathtt{B}_i\times\mathtt{B}_j=-\varepsilon_{ijk}\mathtt{B}_k,
\end{equation}
so that the passive left-rotation is given by
\begin{equation}
\fett{x}'=\overline{R}\starP\fett{x}\starP R
=e_{\starP}^{-\frac{1}{2}\alpha^k\mathtt{B}_k}\starP\fett{x}\starP
e_{\starP}^{\frac{1}{2}\alpha^k\mathtt{B}_k}=x^i
\big(R_i^j\fett{\sigma}_j\big).
\end{equation}
It is clear that the above transformations generalize to arbitrary
multivectors $A(x^i)$ and that such a multivector is invariant
under a composed active and passive transformation \cite{Kaehler}.
The generator of such a composed transformation is then the sum of
the active and passive generators, so that one has infinitesimally
\begin{equation}
\left[L^i+\frac{1}{2}\mathtt{B}_i,A(x^n)\right]_{\starMP}
=\left[L^i,A(x^n)\right]_{\starM}+\mathtt{B}_i\times A(x^n)
=\left[\varepsilon^{ijk}x^j\frac{\hbar}{\iu}\frac{\partial}{\partial
x^k}+\mathtt{B}_i\times\right]A(x^n).
\end{equation}
In the conventional formalism one says that in quantum mechanics
one has to go over from the angular momentum operator $\hat{L}_i$
to the operator $\hat{J}_i$ that includes also a Pauli matrix. In
geometric algebra this follows from the invariance behavior of
multivectors. Moreover the spin structure appears automatically if
one deforms the minimal substituted Hamiltonian with the Moyal
star product as shown in \cite{Deform6}. The star eigenfunctions
of this Hamiltonian are then multivectors \cite{Deform3} that
correspond to the Pauli spinors \cite{Francis1}.

The same argumentation is better known from Dirac theory. A vector
in the Minkowski space with basis vectors $\fett{\gamma}_{\mu}$ is
given by $\fett{x} =x^{\mu}\fett{\gamma}_{\mu}$ and the active
transformations can be done with a four dimensional Moyal star
product
\begin{equation}
\starM=\exp\left[\frac{\iu\hbar}{2}\eta^{\mu\nu}
\left(\frac{\overleftarrow{\partial}}{\partial x^{\mu}}
\frac{\overrightarrow{\partial}}{\partial p^{\nu}}
-\frac{\overleftarrow{\partial}}{\partial p^{\mu}}
\frac{\overrightarrow{\partial}}{\partial x^{\nu}}\right)\right],
\end{equation}
where the nonstandard metric
$\eta_{\mu\nu}=\mathrm{diag}(-1,1,1,1)$ should be chosen. The
generators of an active Lorentz transformation are
\begin{equation}
M^{\mu\nu}=x^{\mu}p^{\nu}-p^{\mu}x^{\nu},
\end{equation}
where the generators of boosts and rotations are
\begin{equation}
K^i=M^{01}\qquad\mathrm{and}\qquad
L^i=\sum_{j<k}\varepsilon^{ijk}M^{jk}.
\end{equation}
They form the following active Moyal star-commutator algebra
\begin{equation}
\left[L^i,L^j\right]_{\starM}=\iu\hbar\varepsilon^{ijk}L^k,\qquad
\left[L^i,K^j\right]_{\starM}=\iu\hbar\varepsilon^{ijk}K^k
\qquad\mathrm{and}\qquad
\left[K^i,K^j\right]_{\starM}=-\iu\hbar\varepsilon^{ijk}L^k,
\label{activeLorentzalg}
\end{equation}
so that an active Lorentz transformation of the four-vector
$\fett{x}=x^{\mu}\fett{\gamma}_{\mu}$ is given by
\begin{equation}
\fett{x}'=e_{\starM}^{-\frac{\iu}{\hbar}\alpha_{\mu\nu}M^{\mu\nu}}
\starM \fett{x}\starM
e_{\starM}^{\frac{\iu}{\hbar}\alpha_{\mu\nu}M^{\mu\nu}}
=\big(\Lambda^{\mu}_{\nu}x^{\nu}\big)\fett{\gamma}_{\mu},
\end{equation}
where $\Lambda_{\nu}^{\mu}$ is the well known Lorentz
transformation matrix.

The corresponding passive Lorentz transformation is generated by
the bivectors
\begin{equation}
\sigma_{\mu\nu}=\frac{I_{(4)}}{2}\starP\left[\fett{\gamma}_{\mu},
\fett{\gamma}_{\nu}\right]_{\starP},
\end{equation}
where $I_{(4)}=\fett{\gamma}_0\fett{\gamma}_1\fett{\gamma}_2
\fett{\gamma}_3$ is the pseudoscalar. The generators for the
passive boosts and rotations are
\begin{equation}
\mathtt{K}_i=\frac{1}{2}\sigma_{0i}\qquad\mathrm{and}\qquad
\mathtt{L}_i=\frac{1}{2}\sum_{j<k}\varepsilon_{ijk}\sigma_{jk}
\end{equation}
and they satisfy in the case of the nonstandard metric (for the
standard metric one has to replace i by -i in the active Lorentz
algebra (\ref{activeLorentzalg}) and $I_{(4)}$ by $-I_{(4)}$ in
the passive Lorentz algebra (\ref{passiveLorentzalg})):
\begin{equation}
\left[\mathtt{L}_i,\mathtt{L}_j\right]_{\starP}
=-I_{(4)}\starP\varepsilon_{ijk}\mathtt{L}_k,\qquad
\left[\mathtt{L}_i,\mathtt{K}_j\right]_{\starP}
=-I_{(4)}\starP\varepsilon_{ijk}\mathtt{K}_k,\quad\mathrm{and}\quad
\left[\mathtt{K}_i,\mathtt{K}_j\right]_{\starP}
=I_{(4)}\starP\varepsilon_{ijk}\mathtt{L}_k.
\label{passiveLorentzalg}
\end{equation}
The passive Lorentz transformation is then given by
\begin{equation}
\fett{x}'=e_{\starP}^{\frac{1}{4}I_{(4)}\starP\alpha^{\mu\nu}\sigma_{\mu\nu}}
\starP\fett{x}\starP
e_{\starP}^{-\frac{1}{4}I_{(4)}\starP\alpha^{\mu\nu}\sigma_{\mu\nu}}
=x^{\mu}\big(\Lambda_{\mu}^{\nu}\fett{\gamma}_{\nu}\big).
\end{equation}
In Dirac theory the passive transformations are constructed a
posteriori by demanding the invariance of the four-vector
$p_{\mu}\fett{\gamma}^{\mu}$, just as the basis vectors of
space-time are discovered a posteriori in a tuple notation by
factorizing the Klein-Gordon equation.

\section{Symplectic Vector Manifolds}
\setcounter{equation}{0}
A symplectic vector space can be considered as a $2d$-dimensional
euclidian space with vectors
\begin{equation}
\fett{z}=z^a\fett{\zeta}_a=q^m\fett{\eta}_m+p^m\fett{\rho}_m,
\label{xphasenraum}
\end{equation}
where $a=1,\ldots,2d$ and $m=1,\ldots,d$, and a closed two-form
\begin{equation}
\Omega=\frac{1}{2}\Omega_{ab}\fett{\zeta}^a\fett{\zeta}^b
=\sum_{m=1}^d \fett{\eta}^m\fett{\rho}^m
=\sum_{m=1}^d\fett{d}q^m\fett{d}p^m,\label{OmegaDef}
\end{equation}
where $\Omega_{ab}$ is a non-degenerate, antisymmetric matrix
\cite{McDuff}. The euclidian metric on the vector space defines a
scalar product and a relation between vectors and one-forms. The
two-form $\Omega$ gives now an additional possibility to establish
such structures, i.e.\ one can define the symplectic scalar
product as
\begin{equation}
\fett{z}\cdotSy\fett{w}\equiv\dot{\iota}_{\mbox{\scriptsize
\boldmath$zw$}}\Omega=(\fett{wz})\cdot\Omega
=\fett{z}\cdot(\Omega\cdot\fett{w})=z^a\Omega_{ab}w^b\label{cdotSy}
\end{equation}
and furthermore one can map with $\Omega$ a vector into a one-form
according to $\fett{z}^{\flat}=\dot{\iota}_{\mbox{\scriptsize
\boldmath$z$}}\Omega =\fett{z}\cdot\Omega$ (the other possibility
used in \cite{Hestenes9} is
$\Omega\cdot\fett{z}=-\fett{z}\cdot\Omega$). The inverse map of a
one-form into a vector can be described with the bivector
\begin{equation}
\mathtt{J}=\frac{1}{2}J^{ab}\fett{\zeta}_a\fett{\zeta}_b
=\frac{1}{2}\sum_{a,b=1}^{2d}\Omega_{ab}\fett{\zeta}_a\fett{\zeta}_b
=\sum_{m=1}^d\fett{\eta}_m\fett{\rho}_m,
\end{equation}
so that the vector corresponding to a one-form $\fett{\omega}$ is
given by $\fett{\omega}^{\natural}=\mathtt{J}\cdot\fett{\omega}$.
The map $\natural$ should be inverse to $\flat$, from which
$J^{ab}=(\Omega_{ab}^{-1})^T=\Omega^{ba}$ follows. Especially with
the nabla operator
$\fett{\nabla}=\fett{d}=\fett{\zeta}^a\partial_a$ and
$\fett{d}^{\natural}=\mathtt{J}\cdot\fett{d}$ the Hamilton
equations can be written as in \cite{Hestenes9}:
\begin{equation}
\dot{\fett{z}}=\fett{d}^{\natural}H.
\end{equation}
Furthermore the Poisson bracket can be written as
\begin{equation}
\{F,G\}_{PB}=F\,\overleftarrow{\fett{d}}\cdotSy\overrightarrow{\fett{d}}\, G
=J^{ab}\frac{\partial F}{\partial x^a}\frac{\partial G}{\partial
x^b}.
\end{equation}
The bivector $\mathtt{J}$ plays the role of the
compatible complex structure \cite{McDuff}, because
\begin{equation}
(\fett{z}\cdot\mathtt{J})\cdotSy(\fett{w}\cdot
\mathtt{J})=\fett{z}\cdotSy\fett{w} \qquad\mathrm{and}\qquad
\fett{z}\cdotSy(\fett{z}\cdot\mathtt{J})>0
\qquad\forall\fett{z}\neq 0.
\end{equation}
Furthermore one has $\mathtt{J}\cdot\mathtt{J}=-1$,
$(\fett{z}\cdot\mathtt{J})\cdot\mathtt{J}=-\fett{z}$ and the
symplectic scalar product can be written as
$\fett{z}\cdotSy\fett{w}=(\fett{z}\cdot\mathtt{J})\cdot\fett{w}$.
A metric space with a two-form $\Omega$ and a compatible complex
structure is a K\"ahler space.

A symplectic vector manifold is an even-dimensional vector
manifold with a closed two-form $\Omega(\fett{x})
=\frac{1}{2}\Omega_{ij}\fett{\xi}^i\fett{\xi}^j$, i.e.\ with
$\partial_i\Omega_{jk}+\partial_j\Omega_{ki}+\partial_k\Omega_{ij}=0$.
The tangent spaces at the symplectic vector manifold are
symplectic vector spaces. A vector field $\fett{z}(\fett{x})$ on a
symplectic vector manifold is symplectic if $\fett{z}^{\flat}$ is
closed, i.e.\ if $\fett{d}(\fett{z}\cdot\Omega)=0$. Symplectic
vector fields conserve the symplectic structure, i.e.\
$\mathscr{L}_{\mbox{\scriptsize \boldmath$z$}}\Omega
=\fett{d}\dot{\iota}_{\mbox{\scriptsize \boldmath$z$}}\Omega=0$
and they form an algebra under the Jacobi-Lie bracket, i.e.\ for
two symplectic vector fields $\fett{z}(\fett{x})$ and
$\fett{w}(\fett{x})$ one has
$\fett{d}\big(\left[\fett{z},\fett{w}\right]_
{JLB}\cdot\Omega\big)=0$. If $\fett{z}^{\flat}$ is not only
closed, but is also exact the vector field is called hamiltonian.
According to the Poincar\'{e} lemma every closed form is locally
exact, so that a symplectic vector field is locally hamiltonian.
This means for a local hamiltonian vector field
$\fett{h}_H$ exists locally a function $H$ so that
\begin{equation}
\fett{h}_H\cdot\Omega=\fett{d}H.\label{hHOmegadH}
\end{equation}
In the coordinate basis the hamiltonian vector field reads
\begin{equation}
\fett{h}_H=\fett{d}^{\natural}H=J^{ij}(\partial_jH)\fett{\xi}_i.
\label{hHDef}
\end{equation}
With a hamiltonian vector field the Poisson bracket can then be
written as
\begin{equation}
\mathscr{L}_{{\mbox{\scriptsize \boldmath$h$}}_H}F
=\fett{h}_H\cdot\fett{d}F=\{F,H\}_{PB},
\end{equation}
or, using (\ref{hHOmegadH}) in this equation, as
\begin{equation}
\{F,G\}_{PB}=\dot{\iota}_{\mbox{\scriptsize{\boldmath$h$}}_F
\mbox{\scriptsize{\boldmath$h$}}_G}\Omega
=(\fett{h}_G\fett{h}_F)\cdot\Omega.\label{OmegaPB}
\end{equation}
It is easy to see that the hamiltonian vector fields form a Lie
subalgebra of the symplectic vector fields with
\begin{equation}
\left[\fett{h}_F,\fett{h}_G\right]_{JLB}=-\fett{h}_{\{F,G\}_{PB}}.
\label{JLBhFhG}
\end{equation}
Given a symplectic vector field $\fett{z}$ that preserves the
Hamilton function $H$, i.e.\
$\mathscr{L}_{\mbox{\scriptsize{\boldmath$z$}}}\Omega
=\Omega\phantom{\fett{0}}$ \!\!\!\!\! and
$\mathscr{L}_{\mbox{\scriptsize{\boldmath$z$}}}H
=0\phantom{\fett{0}}$\!\!\!, this symplectic vector field
$\fett{z}$ can be written locally as a hamiltonian vector field
$\fett{h}_F$ with
\begin{equation}
\mathscr{L}_{\mbox{\scriptsize{\boldmath$h$}}_F}H=\fett{h}_F\cdot
\fett{d}H=\{F,H\}_{PB}=0,
\end{equation}
which shows that $F$ is a conserved quantity. This is Noethers
theorem for the symplectic case.

The metric $g_{ij}(\fett{x})$ on the vector manifold is
induced by the ambient space and exists naturally on the vector
manifold. It was used in the above discussion just to contract vector
fields and forms with the scalar product. But this contraction is
actually independent of the metric. The metric can be used to
define a compatible almost complex structure. This is here a
bivector field $\mathtt{J}(\fett{x})$, that maps via the scalar
product a tangent vector into another tangent vector. If the
structures $g_{ij}(\fett{x})$, $\mathtt{J}(\fett{x})$ and
$\Omega(\fett{x})$ are compatible the metric scalar product of two
tangent vectors $\fett{a}$ and $\fett{b}$ in a point $\fett{x}$
can be written as
$\fett{a}\cdot\fett{b}=\fett{a}\cdotSy(\fett{b}\cdot\mathtt{J})$
and the symplectic product can be written as
$\fett{a}\cdotSy\fett{b}=(\fett{a}\cdot\mathtt{J})\cdot\fett{b}$.
A symplectic vector manifold with these three compatible
structures corresponds to a K\"ahler vector manifold (if the Nijenhuis
torsion vanishes).

Symplectic manifolds of special physical interest are cotangent
bundles, for which the symplectic two-form is globally exact. The
cotangent bundle of a $d$-dimensional euclidian vector space is a
$2d$-dimensional euclidian vector space with elements
$\fett{q}+\fett{\pi}=q^m\fett{\eta}_m+p_m\fett{\rho}^m$. On this
vector space one can define with a vector
$\fett{a}+\fett{\omega}=a^m\fett{\eta}_m+\omega_m\fett{\rho}^m$ a
canonical one-form $\fett{\theta}(\fett{q}+\fett{\pi})$ by
\begin{equation}
(\fett{a}+\fett{\omega})\cdot\fett{\theta}(\fett{q}+\fett{\pi})=a^mp_m,
\label{thetadef}
\end{equation}
so that $\fett{\theta}=p_m\fett{\eta}^m=p_m\fett{d}q^m$, where the
nabla operator is given by $\fett{\nabla}=\fett{d}
=\fett{\eta}^m\frac{\partial}{\partial q^m}
+\fett{\rho}_m\frac{\partial}{\partial p_m}$. The symplectic
two-form on the cotangent bundle can then be obtained as
\begin{equation}
\Omega=-\fett{d}\fett{\theta}=\fett{\eta}^m\fett{\rho}_m
=\fett{d}q^m\fett{d}p_m.
\end{equation}
The above definitions generalize readily to the case of a
cotangent bundle of a $d$-dimensional vector manifold. In a
$(2d+2)$-dimensional ambient vector space with basis vectors
$\fett{\sigma}_a$ and $\fett{\tau}^a$ the cotangent bundle can be
described as the $2d$-dimensional vector manifold
$(\fett{q}+\fett{\pi})(q^i,p_i)=q^a(q^i)\fett{\sigma}_a
+p_j\xi^j_a(q^i)\fett{\tau}^a$, with tangent vectors
$\fett{a}+\fett{\omega}=a^i\xi_i^a\fett{\sigma}_a
+\omega_i\xi_a^i\fett{\tau}^a$. With a projection operator
$T\pi_{\mbox{\scriptsize{\boldmath$q$}}}\phantom{\fett{0}}\!\!\!\!$
defined as
\begin{equation}
T\pi_{\mbox{\scriptsize{\boldmath$q$}}}(\fett{a}+\fett{\omega})
=T\pi_{\mbox{\scriptsize{\boldmath$q$}}}
(\fett{a})=a^i\xi_i^a\fett{\tau}_a,\label{Tpiqdef}
\end{equation}
one can write (\ref{thetadef}) as
$(\fett{a}+\fett{\omega})\cdot\fett{\theta}(\fett{q}+\fett{\pi})
=T\pi_{\mbox{\scriptsize{\boldmath$q$}}}(\fett{a}+\fett{\omega})
\cdot\fett{\pi}$, so that
$\fett{\theta}=p_i\fett{\xi}^i=p_i\fett{d}q^i$.

In the discussion so far the symplectic structure was defined via
a two-form and a compatible metric star product, which led to a
K\"ahler vector manifold. But a metric structure is actually not
necessary to define a symplectic structure. In the case of a
cotangent bundle it suffices to use the natural duality on this
space. This duality can also be described with a star product, for
example on the cotangent bundle of a vector space one can define
\begin{equation}
F\starD G=F\,\exp\left[
\frac{\overleftarrow{\partial}}{\partial\fett{\eta}_a}
\frac{\overrightarrow{\partial}}{\partial\fett{\rho}^a}\right]\,G,
\end{equation}
so that (\ref{thetadef}) reads
$\dot{\iota}_{\mbox{\scriptsize{\boldmath$a$}}+
\mbox{\scriptsize{\boldmath$\omega$}}}\fett{\theta}(\fett{q}+\fett{\pi})
=\dot{\iota}_{\mbox{\scriptsize{\boldmath$a$}}}\fett{\pi}
=\langle\fett{a}\starD\fett{\pi}\rangle_0=\fett{a}\cdotD\fett{\pi}$
and further $\dot{\iota}_{(\mbox{\scriptsize{\boldmath$a$}}
+\mbox{\scriptsize{\boldmath$\omega$}})(\mbox{\scriptsize{\boldmath$b$}}
+\mbox{\scriptsize{\boldmath$\chi$}})}
\Omega=\fett{a}\cdotD\fett{\chi}-\fett{b}\cdotD\fett{\omega}$,
which can easily be generalized to manifolds \cite{Marsden}. The
other possibility is to define a symplectic star product, by using
$\Omega_{ij}$ instead of the metric $\eta_{ij}$ in the fermionic
star product. On a $2d$-dimensional vector space the symplectic
star product in Darboux coordinates is given by
\begin{equation}
F\starSy G=F\,\exp\left[\Omega_{ab}
\frac{\overleftarrow{\partial}}{\partial\fett{\zeta}_a}
\frac{\overrightarrow{\partial}}{\partial\fett{\zeta}_b}\right]\,G
=F\,\exp\left[\sum_{m=1}^{d}\left(
\frac{\overleftarrow{\partial}}{\partial\fett{\eta}_m}
\frac{\overrightarrow{\partial}}{\partial\fett{\rho}_m}
-\frac{\overleftarrow{\partial}}{\partial\fett{\rho}_m}
\frac{\overrightarrow{\partial}}{\partial\fett{\eta}_m}\right) \right]\,G.
\end{equation}
On a $2d$-dimensional vector manifold the tangent space can also
be spanned by Darboux basis vectors
$\fett{\eta}_m=\eta_m^i\fett{\xi}_i$ and $\fett{\rho}_m
=\rho_m^i\fett{\xi}_i$ so that one has analogously
\begin{equation}
F\starSy G=F\,\exp\left[\Omega_{ij}
\frac{\overleftarrow{\partial}}{\partial\fett{\xi}_i}
\frac{\overrightarrow{\partial}}{\partial\fett{\xi}_j}\right]\,G
=F\,\exp\left[\sum_{m=1}^{d}\left(
\frac{\overleftarrow{\partial}}{\partial\fett{\eta}_m}
\frac{\overrightarrow{\partial}}{\partial\fett{\rho}_m}
-\frac{\overleftarrow{\partial}}{\partial\fett{\rho}_m}
\frac{\overrightarrow{\partial}}{\partial\fett{\eta}_m}\right) \right]\,G.
\end{equation}
The indices are now lowered and raised with $\Omega_{ij}$, i.e.\
for a tangent vector $\fett{a}=a^i\fett{\xi}_i$ one has
$a_i=\Omega_{ij}a^j$ and $\fett{\xi}^i=\Omega^{ij}\fett{\xi}_j$,
where $\Omega_{ij}\Omega^{jk}=\delta_i^k$. The relations $\flat$
and $\natural$ between vectors and one-forms can then be written
as $\fett{a}^{\flat}
=a^i\Omega_{ij}\fett{\xi}^j=(\Omega_{ji}^Ta^i)\fett{\xi}^j$ and
$\fett{\omega}^{\natural}=\omega_i\Omega^{ij}\fett{\xi}_j
=(\Omega^{ji\,T}\omega_i)\fett{\xi}_j
=J^{ji}\omega_i\fett{\xi}_j$. Furthermore it follows for the scalar
products
\begin{equation}
\fett{\xi}_i\cdotSy\fett{\xi}_j=\Omega_{ij},\qquad
\fett{\xi}^i\cdotSy\fett{\xi}_j=-\fett{\xi}_j\cdotSy\fett{\xi}^i
=\delta_j^i\qquad\mathrm{and}\qquad
\fett{\xi}^i\cdotSy\fett{\xi}^j=-\Omega^{ij}=J^{ij}.
\end{equation}
If one establishes the symplectic structure with the symplectic
star product and not with a metric star product and a two-form,
the contraction of vectors and one-forms has to be defined with
the symplectic scalar product
$\fett{\xi}_i\cdotSy\fett{\xi}^j=-\delta_i^j$. This leads to a
different sign structure compared with the case of a metric star
product, for example instead of (\ref{hHOmegadH}) one has for a
hamiltonian vector field on a vector space with a symplectic star
product $\fett{h}_H\cdotSy\Omega=-\fett{d}H$ and since
$\fett{a}\cdotSy\fett{\partial} =-\fett{a}\cdot\fett{\partial}$
there is no minus sign on the right side of (\ref{JLBhFhG}). So
these two sign conventions correspond to the use of a metric or
a symplectic star product on the vector space.

\section{Active and passive Transformations on the Phase Space}
\setcounter{equation}{0}
A flat phase space can be considered as an $2d$-dimensional
euclidian vector space with vectors (\ref{xphasenraum}) and a
two-form (\ref{OmegaDef}). The time development is described by
the hamiltonian vector field
$\fett{h}_H=\dot{q}^n\fett{\eta}_n+\dot{p}^n\fett{\rho}_n
=J^{ij}\partial_j H\fett{\zeta}_i$, so that one has for a scalar
phase space function $f$
\begin{equation}
\dot{f}=\dot{\fett{z}}\cdot(\fett{d}f)=(\fett{h}_H\cdot
\fett{d})f=\mathscr{L}_{\mbox{\scriptsize\boldmath$h$}_H}f=
\{f,H\}_{PB}.
\end{equation}
where $\fett{h}_H\cdot\fett{d}$ is the Liouville operator. The
above equation for the time development can immediately be
generalized from 0-forms $f$ to arbitrary $r$-forms. For example
the time development of the symplectic two-form is given by
$\dot{\Omega}=\mathscr{L}_{\mbox{\scriptsize\boldmath$h$}_H}\Omega=0$,
which means that the symplectic form is preserved by the time
evolution.

The temporal development of a system can be described by an active
time transformation of the coefficients, which corresponds to the
Hamilton equations
\begin{equation}
\dot{z}^i=\mathscr{L}_{\mbox{\scriptsize\boldmath$h$}_H}z^i
=J^{ij}\partial_jH\label{xiHamil2}.
\end{equation}
In the formalism of geometric algebra it is also possible to write
down a time transformation of the basis vectors
\begin{equation}
\dot{\fett{\zeta}}_i=\mathscr{L}_{\mbox{\scriptsize\boldmath$h$}_H}
\fett{\zeta}_i=-J^{jk}\partial_k\partial_iH\fett{\zeta}_j,
\label{pasHamil}
\end{equation}
which corresponds to the Jacobi equation that appeared in the path
integral formulation of classical mechanics \cite{Gozzi1}.

Active and passive time development can directly be discussed for
the example of the harmonic oscillator. The Hamiltonian
$H=\frac{1}{2}(p^2+q^2)$ generates via the star exponential
$U(t)=e_{\starM}^{-\frac{i}{\hbar}Ht}$ an active rotation of the
state vector $\fett{z}_0=q\fett{\eta}+p\fett{\rho}$ according to
\cite{Zachos3}
\begin{equation}
\fett{z}(t)=\overline{U(t)}\starM \fett{z}_0\starM U(t)=(q \cos t+
p\sin t)\fett{\eta}+(-q\sin t+p\cos t)\fett{\rho}
=q(t)\fett{\eta}+p(t)\fett{\rho}.
\end{equation}
The same transformation passively can be achieved with the rotor
$R(t)=e_{\starP}^{\frac{1}{2} \mathtt{H}t}$ and the bivector
$\mathtt{H}=\fett{\eta\rho}$ as
\begin{equation}
\fett{z}(t)=R(t)\starP\fett{z}_0\starP\overline{R(t)}= q(\cos
t\,\fett{\eta}-\sin t\,\fett{\rho})+p(\sin t\,\fett{\eta}+\cos
t\,\fett{\rho})=q\fett{\eta}(t)+p\fett{\rho}(t).
\end{equation}
With the hamiltonian vector-field $\fett{h}_H
=p\fett{\eta}-q\fett{\rho}$ and the relation $\{f,g\}_{PB}
=\lim_{\hbar\rightarrow 0}\frac{1}{\iu\hbar}
\left[f,g\right]_{\starM}$ the active Hamilton equations
$\dot{z}^i =\mathscr{L}_{\mbox{\scriptsize\boldmath$h$}_H}z^i$ can
be written as
\begin{equation}
\dot{q}=\lim_{\hbar\rightarrow 0}\frac{1}{\iu\hbar}
\left[q,H\right]_{\starM}=p\qquad\mathrm{and}\qquad
\dot{p}=\lim_{\hbar\rightarrow 0}\frac{1}{\iu\hbar}
\left[p,H\right]_{\starM}=-q.
\end{equation}
With (\ref{pasHamil}) one can then calculate the corresponding
time inverted passive Hamilton equations. Using the Clifford star
commutator defined by
\begin{equation}
\left[A_{(r)},B_{(s)}\right]_{\starP}=A_{(r)}\starP
B_{(s)}-(-1)^{rs} B_{(s)}\starP A_{(r)}\label{Cliffstarcommu}
\end{equation}
these equations can be written as
\begin{equation}
\dot{\fett{\eta}}=\frac{1}{\iu}\left[\fett{\eta},\mathtt{H}\right]_{\starP}
=\fett{\rho}\qquad\mathrm{and}\qquad
\dot{\fett{\rho}}=\frac{1}{\iu}\left[\fett{\rho},\mathtt{H}\right]_{\starP}
=-\fett{\eta},
\end{equation}
where $\mathtt{H} =\frac{\iu}{2}\fett{\eta\rho}$ is the passive
Hamiltonian. The passive Hamiltonian is connected with the active
one over (\ref{Cliffstarcommu}) and (\ref{pasHamil}) by
\begin{equation}
\frac{1}{\iu} \left[\fett{\zeta}_i,\mathtt{H}\right]_{\starP}
=-J^{jk}\partial_k\partial_i H\fett{\zeta}_j.\label{actipassiHami}
\end{equation}
The passive Hamiltonian $\mathtt{H}$ is here just the free
Hamiltonian of pseudoclassical mechanics \cite{Berezin1} (the
additional factor $\frac{1}{2}$ is due to the definition of the
Clifford product which is defined without a factor $\frac{1}{2}$,
see for example \cite{Deform6}).

A Lagrangian that takes into account both the time development
according to (\ref{xiHamil2}) and the time development according
to (\ref{pasHamil}) should be called the extended Lagrangian and has
the form
\begin{eqnarray}
\widetilde{{\cal{L}}}_E&=&y_i\left(\dot{z}^i-J^{ij}\partial_jH
\right)+\iu\fett{\zeta}_j\left(\partial_t\delta_l^j-
J^{jk}\partial_l\partial_kH\right)\fett{\lambda}^l
\nonumber\\
&=&y_i\dot{z}^i+\iu\fett{\zeta}_j\dot{\fett{\lambda}}^j-
\widetilde{{\cal{H}}}_E,
\end{eqnarray}
where the extended Hamiltonian $\widetilde{{\cal{H}}}_E$ is given
by
\begin{equation}
\widetilde{{\cal{H}}}_E=y_iJ^{ij}\partial_jH
+\iu\fett{\zeta}_jJ^{jk}\partial_l\partial_kH\fett{\lambda}^l.
\end{equation}

The extended Lagrangian first appeared in the path integral
approach to classical mechanics \cite{Gozzi1,Gozzi2}, where the
classical analogue of the quantum generating functional was
considered:
\begin{equation}
Z_{CM}\left[J\right]=N\int Dz\,\delta\left[z(t)-z_{cl}(t)\right]
\;\exp\left[\int dt\, J\phi\right].
\end{equation}
The delta function here constrains all possible trajectories to
the classical trajectory obeying (\ref{xiHamil2}). It can be
written as
\begin{equation}
\delta\left[z(t)-z_{cl}(t)\right]=\delta\left[\dot{z}^i-\Omega^{ij}
\partial_{j}H\right]\det\left[\delta^i_j\partial_t-\Omega^{ik}
\partial_k\partial_jH\right].
\end{equation}
The delta function on the right side can be expressed by a Fourier
transform
\begin{equation}
\delta\left[\dot{z}^i-\Omega^{ij}\partial_jH\right] =\int
Dy_i\exp\left[\iu\int dt\,y_i
\left(\dot{z}^i-\Omega^{ij}\partial_jH\right)\right]
\end{equation}
and the determinant can be written in terms of Grassmann variables
as
\begin{equation}
\det\left[\delta^i_j\partial_t-\Omega^{ik}\partial_k\partial_jH\right]
=\int D\fett{\lambda}^iD\fett{\zeta}_i\,\exp\left[-\int
dt\,\fett{\zeta}_i
\left[\delta^i_j\partial_t-\Omega^{ik}\partial_k\partial_jH\right]
\fett{\lambda}^j\right],
\end{equation}
so that $Z_{CM}\left[0\right]$ becomes
\begin{equation}
Z_{CM}\left[0\right]=\int
Dz^iDy_iD\fett{\lambda}^jD\fett{\zeta}_j\, \exp\left[\iu\int
dt\,\widetilde{{\cal{L}}}_E\right].
\end{equation}
The important point is here, that the path integral formalism of
classical mechanics gives the fermionic basis vectors of geometric
algebra the physical interpretation of ghosts. On the other hand
the superanalytic formulation of geometric algebra has naturally
the fermionic structures that in the conventional formalism have
to be added ad hoc and per hand.

The $z^i$ and $\fett{\zeta}_i$ form together with the newly
introduced variables $y_i$ and $\fett{\lambda}^i$ the extended
phase space. On this extended phase space one can then introduce
an extended canonical structure. This can easily be done in
analogy to the Moyal and the Clifford star product structures of
the phase space. Defining the extended Moyal-Clifford star product
as
\begin{equation}
F\starEMC G=F\,\exp\left[\frac{\iu}{2}\left(\frac{\overleftarrow{\partial}}
{\partial z^k}\frac{\overrightarrow{\partial}}{\partial y_k}
-\frac{\overleftarrow{\partial}}{\partial
y_k}\frac{\overrightarrow{\partial}}{\partial z^k} \right)
+\frac{1}{2}\left(\frac{\overleftarrow{\partial}}
{\partial\fett{\lambda}^k}\frac{\overrightarrow{\partial}}{\partial\fett{\zeta}_k}+
\frac{\overleftarrow{\partial}}{\partial\fett{\zeta}_k}
\frac{\overrightarrow{\partial}}{\partial\fett{\lambda}^k}\right)\right]\,G
\end{equation}
the extended Poisson bracket has the form
\begin{equation}
\{F,G\}_{EPB}=\frac{1}{\iu}\left[F\starEMC G -(-1)^{\epsilon
(F)\epsilon (G)}G\starEMC F\right], \label{EPBDef}
\end{equation}
where $\epsilon (F)$ gives the Grassmann grade of $F$. In the
bosonic part of the extended Clifford star product a factor
$\hbar$ can be included like in the Moyal product, so that in the
definition of the extended Poisson bracket (\ref{EPBDef}) the
limit $\hbar\rightarrow 0$ has to be taken. The extended canonical
relations are then given by
\begin{equation}
\{z^i,y_j\}_{EPB}=\delta_j^i\qquad\mathrm{and}\qquad
\{\fett{\zeta}_i,\fett{\lambda}^j\}_{EPB}=-\iu\delta_i^j,
\end{equation}
while all other extended Poisson brackets vanish. Furthermore one
can calculate the equations of motion as
\begin{eqnarray}
\dot{z}^i&=&\{z^i,\widetilde{{\cal{H}}}_E\}_{EPB}
=\Omega^{ij}\partial_jH,\\
\dot{\fett{\zeta}}_i&=&\{\fett{\zeta}_i,\widetilde{{\cal{H}}}_E\}_{EPB}
=-\Omega^{jk}\partial_k\partial_iH\fett{\zeta}_j,\\
\dot{y}_i&=&\{y_i,\widetilde{{\cal{H}}}_E\}_{EPB}
=-z_j\Omega^{jk}\partial_k\partial_iH
-\iu\fett{\zeta}_j\Omega^{jk}\partial_k\partial_l\partial_iH
\fett{\lambda}^l,\\
\dot{\fett{\lambda}}^i&=&\{\fett{\lambda}^i,\widetilde{{\cal{H}}}_E\}_{EPB}
=\Omega^{ij}\partial_j\partial_kH\fett{\lambda}^k.
\end{eqnarray}
The extended Hamiltonian also generates the time development of
$r$-vectors and $r$-forms according to \cite{Gozzi5}
\begin{equation}
\dot{X}=\mathscr{L}_{\mbox{\scriptsize \boldmath$h$}}X=\{X,
\widetilde{{\cal{H}}}_E\}_{EPB}.
\end{equation}

Having now a superanalytic formalism for classical mechanics that
takes into account active and passive time development, one can
ask if there is a supersymmetry in this formalism, i.e. a symmetry
that relates the bosonic coefficients with the fermionic basis
vectors. This supersymmetry was found by Gozzi et al. in
\cite{Gozzi1}. There was shown that $\widetilde{{\cal{H}}}_E$ is
invariant under the following BRST-transformation
\begin{equation}
\delta z^k=\fett{\varepsilon}\fett{\lambda}^k ,\qquad \delta
\fett{\zeta}_k=\iu\fett{\varepsilon}y_k,\qquad
\delta\fett{\lambda}^k=\delta y_k=0
\end{equation}
and the following anti-BRST-transformation
\begin{equation}
\delta z^k=-\fett{\varepsilon}\Omega^{kl}\fett{\zeta}_l ,\qquad
\delta\fett{\lambda}^k=\iu\overline{\fett{\varepsilon}}\Omega^{kl}y_l,\qquad
\delta\fett{\zeta}_k=\delta y_k=0,
\end{equation}
where $\fett{\varepsilon}$ and $\overline{\fett{\varepsilon}}$ are
Grassmann variables. These symmetries are generated by
\begin{equation}
\fett{Q}_{BRST}=y_j\fett{\lambda}^j\qquad\mathrm{and}\qquad
\fett{Q}_{\overline{BRST}}=\fett{\zeta}_j\Omega^{jk}y_k
\end{equation}
according to $\delta X=\{X,\fett{\varepsilon}\fett{Q}_{BRST}+
\overline{\fett{\varepsilon}}\fett{Q}_{\overline{BRST}}\}_{EPB}.$
The two charges $\fett{Q}_{BRST}$ and $\fett{Q}_{\overline{BRST}}$
are conserved, i.e.
\begin{equation}
\{\fett{Q}_{BRST},\widetilde{{\cal{H}}}_E\}_{EPB}=
\{\fett{Q}_{\overline{BRST}},\widetilde{{\cal{H}}}_E\}_{EPB}=0
\end{equation}
and fulfill
\begin{equation}
\{\fett{Q}_{BRST},\fett{Q}_{BRST}\}_{EPB}=
\{\fett{Q}_{\overline{BRST}},\fett{Q}_{\overline{BRST}}\}_{EPB}=
\{\fett{Q}_{BRST},\fett{Q}_{\overline{BRST}}\}_{EPB}=0.
\end{equation}

\section{Poisson Vector Manifolds}
\setcounter{equation}{0}
A vector manifold $M$ with a bivector $\mathtt{J}(\fett{x})
=\frac{1}{2}J^{ij}\fett{\xi}_i \fett{\xi}_j$ and
\begin{equation}
J^{ij}\partial_iJ^{kl}+J^{ik}\partial_iJ^{lj}+J^{il}\partial_iJ^{jk}=0
\label{JJacobi}
\end{equation}
is a Poisson vector manifold, where (\ref{JJacobi}) can be written
with (\ref{SNBDef}) as
$\left[\mathtt{J},\mathtt{J}\right]_{SNB}=0$. The bivector
$\mathtt{J}$ defines as discussed above a map from
$T^*$$_{\!\!\!\!{\mbox{\scriptsize{\boldmath$x$}}}}$$M$ to
$T$$_{\!\!\mbox{\scriptsize{\boldmath$x$}}}$$M$ by
$\fett{\alpha}^{\natural}=\mathtt{J}\cdot\fett{\alpha}
=J^{ij}\alpha_j\fett{\xi}_i$, where
$\fett{\alpha}=\alpha_i\fett{\xi}^i$ is an element of
$T^*$$_{\!\!\!\!\mbox{\scriptsize{\boldmath$x$}}}$$M$. Especially
the hamiltonian vector field (\ref{hHDef}) can be expressed as
\begin{equation}
\fett{h}_H=\dot{\iota}_{\mbox{\scriptsize{\boldmath$d$}}H}\mathtt{J}
=\mathtt{J}\cdot\fett{d}H \label{hHJdH}
\end{equation}
and the Poisson bracket as
\begin{equation}
\{F,G\}_{PB}=\dot{\iota}_{\mbox{\scriptsize{\boldmath$d$}}F
\mbox{\scriptsize{\boldmath$d$}}G}\mathtt{J}
=(\fett{d}G\fett{d}F)\cdot\mathtt{J}, \label{JPB}
\end{equation}
so that the hamiltonian vector field $\fett{h}_H$ can be defined
for all scalar functions $F$ as
\begin{equation}
\fett{h}_H\cdot\fett{d}F=\{F,H\}_{PB}.\label{hHdFFHPB}
\end{equation}
Equating (\ref{OmegaPB}) and (\ref{JPB}) shows how $\Omega$ and
$\mathtt{J}$ determine each other:
\begin{equation}
(\fett{h}_G\fett{h}_F)\cdot\Omega
=(\fett{d}G\fett{d}F)\cdot\mathtt{J}.
\end{equation}
Since a Poisson manifold can be odd-dimensional, the hamiltonian
vector fields do not span in general the tangent space of the
Poisson manifold. This suggests to define the range
$\mathrm{ran}(\mathtt{J}(\fett{x}))$ of $\mathtt{J}(\fett{x})$ as
the span of all tangent vectors that can be expressed as
$\fett{\alpha}^{\natural}$ for a one-form $\fett{\alpha}\in$
$T^*$$_{\!\!\!\!\mbox{\scriptsize{\boldmath$x$}}}$$M$. The range
of $\mathtt{J}(\fett{x})$ is also the span of all hamiltonian
vector fields at $\fett{x}$. The dimension of
$\mathrm{ran}(\mathtt{J}(\fett{x}))$ is the rank of the Poisson
manifold in $\fett{x}$ and equal to the rank of the matrix
$J^{ij}$, which is an even number because of the anti-symmetry of
$J^{ij}$. The even-dimensional vector space
$\mathrm{ran}(\mathtt{J}(\fett{x}))$ is then the tangent space of
a symplectic leaf in the point $\fett{x}$. The Poisson manifold is
foliated by these symplectic leafs. Only when the rank of a
Poisson manifold $M$ is everywhere equal to $\mathrm{dim}(M)$ the
Poisson manifold itself is a symplectic manifold.

The formalism developed so far can now directly be generalized to
multivectors, which leads to Poisson calculus (see \cite{Vaisman}
and the references therein). The $r$-vector that corresponds to an
$r$-form is given by
\begin{equation}
\big(A^{(r)}\big)^{\natural}=\frac{1}{r!}J^{k_1i_1}\ldots
J^{k_ri_i}A_{i_1\ldots i_r}\fett{\xi}_{k_1}\ldots\fett{\xi}_{k_r}
\end{equation}
and in analogy to (\ref{iArBs}) one has $\dot{\iota}_{A^{(r)}}
B_{(s)}=\overline{A^{(r)}}\cdot B_{(s)}$, so that
\begin{equation}
\overline{\fett{\alpha}_1\ldots\fett{\alpha}_r}\cdot\big(A^{(r)}
\big)^{\natural}=(-1)^r\overline{\fett{\alpha}_1^{\natural}\ldots
\fett{\alpha}_r^{\natural}}\cdot A^{(r)}.
\end{equation}
It is then also possible to define a Poisson bracket for one-forms
by
\begin{equation}
\{\fett{\alpha},\fett{\beta}\}_{PB}
=\mathscr{L}_{\mbox{\scriptsize\boldmath$\alpha$}^{\natural}}\fett{\beta}
-\mathscr{L}_{\mbox{\scriptsize\boldmath$\beta$}^{\natural}}\fett{\alpha}
+\fett{d}\big((\fett{\beta\alpha})\cdot\mathtt{J}\big),
\end{equation}
so that $\{\fett{\alpha},\fett{\beta}\}_{PB}^{\natural}
=\big[\fett{\alpha}^{\natural},\fett{\beta}^{\natural}\big]_{JLB}$.
With this Poisson bracket one can further define an exterior
differential $\tilde{\fett{d}}$ in analogy to (\ref{iAr}) as
\begin{eqnarray}
\left(\overline{\fett{\alpha}_1\fett{\alpha}_2\ldots\fett{\alpha}_{r+1}}\right)\cdot
\tilde{\fett{d}}A_{(r)}
&=&\sum_{n=1}^{r+1}(-1)^{n+1}(\fett{\alpha}_n^{\natural}\cdot\fett{\partial})
\left(\overline{\fett{\alpha}_1\ldots\check{\fett{\alpha}}_n\ldots\fett{\alpha}_{r+1}}\right)
\cdot A_{(r)}\nonumber\\
&&+\sum_{m<n}(-1)^{m+n}\left(\overline{\{\fett{\alpha}_m,\fett{\alpha}_n\}_{PB}
\fett{\alpha}_1\ldots\check{\fett{\alpha}}_m\ldots\check{\fett{\alpha}}_n
\ldots\fett{\alpha}_{r+1}}\right)\cdot A_{(r)},
\end{eqnarray}
which can also be written as $\tilde{\fett{d}}A_{(r)}
=\left[\mathtt{J},A_{(r)}\right]_{SNB}$.

The easiest non-constant Poisson tensor fulfilling (\ref{JJacobi})
is a linear tensor
\begin{equation}
J^{ij}(\fett{x})=C^{ij}_kx^k,
\end{equation}
where the antisymmetry of $J^{ij}$ and (\ref{JJacobi}) ensure that
the $C^{ij}_k$ are structure constants of a Lie algebra. The
corresponding Poisson bracket is the so called Lie-Poisson bracket
\begin{equation}
\{F,G\}_{LPB}=C_k^{ij}x^k\partial_i F\partial_jG.
\end{equation}
The most fundamental example is the Lie-Poisson structure on
$\mathfrak{g}^*$. For this purpose one considers the bivector space
spanned by the basis bivectors $\mathtt{B}_i$ with bivector
algebra (\ref{bivecalgebra}) and its reciprocal basis with
two-forms $\Theta^i$, i.e.\ $\overline{\mathtt{B}_i}\cdot\Theta^j
=\delta_i^j$. For scalar-valued functions $F$ and $G$ of general
two-forms $\Theta=\theta_i\Theta^i$ a Lie-Poisson bracket is given
by
\begin{equation}
\{F,G\}_{LPB}(\Theta)=C_{ij}^k\theta_k\frac{\partial F}{\partial
\theta_i}\frac{\partial G}{\partial \theta_j}
=\overline{(\mathtt{d}F\times\mathtt{d}G)}\cdot\Theta,
\end{equation}
where $\mathtt{d}$ is the exterior differential in the bivector
basis: $\mathtt{d}=\mathtt{B}_i\frac{\partial}{\partial
\theta_i}$. In the $SO(3)$-case, where $\Theta^i=\mathtt{B}_i$ the
Lie-Poisson bracket can be written as
\begin{equation}
\{F,G\}_{LPB}(\mathtt{B})=\overline{\mathtt{B}}
\cdot\left((I_{(3)}\starP\fett{d})F\times
(I_{(3)}\starP\fett{d})G\right)=\overline{\mathtt{B}}\cdot
\left(\mathtt{d}F\times\mathtt{d}G\right).\label{SO3LPB}
\end{equation}

The symplectic leafs induced by the symplectic foliation with the
Lie-Poisson bracket on $\mathfrak{g}^*$ are the orbits of the
coadjoint action of the corresponding group $G$ on
$\mathfrak{g}^*$. This can be seen if one considers a scalar
linear function $H(\Theta)=\overline{\mathtt{B}}\cdot\Theta
=b^i\theta_i$ on $\mathfrak{g}^*$ with $\mathtt{d}H=\mathtt{B}$.
For the Lie-Poisson bracket one has then for any scalar function
$F$ on $\mathfrak{g}^*$:
\begin{equation}
\{F,H\}_{LPB}(\Theta)=\overline{(\mathtt{d}F\times\mathtt{d}H)}\cdot\Theta
=-\overline{(\mathtt{B}\times\mathtt{d}F)}\cdot\Theta
=-\overline{(\mathrm{ad}_{\mathtt{B}}\mathtt{d}F)}\cdot\Theta
=-\overline{\mathtt{d}F}\cdot\mathrm{ad}_{\mathtt{B}}^*\Theta.
\end{equation}
On the other hand one can define in analogy to (\ref{hHdFFHPB})
the hamiltonian bivector field $\mathtt{h}_H$ of the Hamilton
function $H(\Theta)$ as
\begin{equation}
\overline{\mathtt{h}_H(\Theta)}\cdot\mathtt{d}F=\{F,H\}_{LPB}(\Theta)
=\overline{(\mathtt{d}F\times\mathtt{d}H)}\cdot\Theta
=-\overline{\mathrm{ad}^*_{\mathtt{B}}\Theta}\cdot\mathtt{d}F,
\end{equation}
so that $\mathtt{h}_H(\Theta)=-\mathrm{ad}_{\mathtt{d}H}^*\Theta$.
This means that the hamiltonian bivector fields $\mathtt{h}_H$
that span the tangent space of the symplectic leaf are, up to a
sign, the generators of the coadjoint action determined by
$\mathtt{B}$. If $\Theta$ varies now over the coadjoint orbit one
can define a skew-symmetric bilinear form on the orbit by
\begin{equation}
\Omega_{\Theta}(\mathrm{ad}_{\mathtt{A}}^*\Theta
,\mathrm{ad}_{\mathtt{B}}^*\Theta)
=\overline{\mathtt{A}\times\mathtt{B}}\cdot\Theta,
\end{equation}
which defines on the coadjoint orbit a symplectic structure, that
is the restriction of the Lie-Poisson bracket to the orbit
\cite{Marsden}. $\Omega_{\Theta}$ can be seen as a generalized
antisymmetric tensor of the form (\ref{generalizedtensor}) that
maps two bivectors into a scalar.

Of special interest is the hamiltonian action of a rotor on a
Poisson vector manifold. The aim is to find the Hamilton function
$P_{\mathtt{B}}$ of the vector field $\mathtt{B}\cdot\fett{x}$,
that is induced according to (\ref{RxRt0}) by the rotor
left-action with bivector $\mathtt{B}$, i.e.
\begin{equation}
\fett{h}_{P_{\mathtt{B}}}=\mathtt{B}\cdot\fett{x}.\label{hFBxB}
\end{equation}
Since $\fett{h}_{P_{\mathtt{B}}}\cdot\fett{d}H
=\{H,P_{\mathtt{B}}\}_{PB}$, it is possible to write the defining
relation for $P_{\mathtt{B}}$ as
\begin{equation}
\{H,P_{\mathtt{B}}\}_{PB}=(\mathtt{B}\cdot\fett{x})
\cdot\fett{d}H,\label{HFBPB}
\end{equation}
for all scalar functions $H$. Furthermore one has for two
bivectors $\mathtt{A}$ and $\mathtt{B}$ with (\ref{JLBhFhG}) and
(\ref{xAxBJLBxAB})
\begin{equation}
\fett{h}_{\{P_{\mathtt{A}},P_{\mathtt{B}}\}_{PB}}
=\fett{h}_{P_{\mathtt{A}\times\mathtt{B}}}.
\end{equation}
While in the symplectic case a symplectic vector field is always
locally hamiltonian, in the Poisson case an infinitesimal Poisson
automorphism is in general not locally hamiltonian. This means
that if the rotor left-action is canonical, i.e.
$\mathscr{L}_{\mathtt{B}\cdot\mbox{\scriptsize{\boldmath$x$}}}\mathtt{J}
=0\phantom{\fett{0}}\!\!\!$, there does not exist in general a
function $P_{\mathtt{B}}$, that fulfills (\ref{hFBxB}). The
additional condition that $\mathtt{B}\cdot\fett{x}$ is also
hamiltonian can be expressed with the momentum map. A momentum map
is here a two-form $\Pi(\fett{x})$ with
\begin{equation}
\dot{\iota}_{\mathtt{B}}\Pi=\overline{\mathtt{B}}\cdot\Pi
=P_{\mathtt{B}}.\label{momentummap}
\end{equation}
So if the hamiltonian vector field $\fett{h}_{P_\mathtt{B}}$
corresponding to the function
$P_{\mathtt{B}}=\overline{\mathtt{B}}\cdot\Pi$ is the same as the
vector field $\mathtt{B}\cdot\fett{x}$ induced by the rotor
left-action, i.e.\ if one has $\fett{h}_{\overline{\mathtt{B}}
\cdot\Pi}=(\mathtt{J}\cdot\fett{d})\cdot(\overline{\mathtt{B}}\cdot\Pi)
=\mathtt{B}\cdot\fett{x}$, then $\Pi$ is a momentum map. If a
momentum map of a rotor action exists and $H$ is a Hamilton
function that is invariant under the rotor action, then equation
(\ref{HFBPB}) reduces to $\{H,P_{\mathtt{B}}\}_{PB}=0$ and the
momentum map is a constant of the motion described by $H$. This
follows because $\{H,P_{\mathtt{B}}\}_{PB}=0$ means that
$P_{\mathtt{B}}$ is constant along the hamiltonian flow of $H$,
which must then also be true for the left hand side of
(\ref{momentummap}), i.e.\ for $\Pi$, because $\mathtt{B}$ is
constant. This is the Noether theorem in the Poisson case.

If a hamiltonian action of a rotor group on a Poisson vector
manifold is given, there are scalar functions $P_{\mathtt{B}_i}$
on the Poisson manifold with $\{P_i,P_j\}_{PB}=-C_{ij}^kP_k$ and
$\left[\fett{h}_{P_i},\fett{h}_{P_j}\right]_{JLB}=C_{ij}^k
\fett{h}_{P_k}$, that generate the hamiltonian action. The
momentum map is then
\begin{equation}
\Pi(\fett{x})=P_{\mathtt{B}_i}(\fett{x})\Theta^i.\label{ThetaFBi}
\end{equation}
A momentum map $\Pi(\fett{x})$ that is determined by a hamiltonian
group action is equivariant, i.e.\ it respects the rotor
left-action on the vector manifold:
\begin{equation}
\Pi(R\starP\fett{x}\starP\overline{R})
=R\starP\Pi(\fett{x})\starP\overline{R},\label{equivarianz}
\end{equation}
which can also be written as
\begin{equation}
\overline{\mathrm{Ad}_R\mathtt{B}}\cdot\Pi(R\starP\fett{x}
\starP\overline{R})\equiv
P_{\mathrm{Ad}_R\mathtt{B}}(R\starP\fett{x}\starP\overline{R})
=P_{\mathtt{B}}(\fett{x})\equiv\overline{\mathtt{B}}\cdot\Pi
(\fett{x}),
\end{equation}
To see that the momentum map (\ref{ThetaFBi}) is equivariant, it
suffices to show the infinitesimal version of (\ref{equivarianz}),
namely $\big(\fett{h}_{P_{\mathtt{B}_j}}\cdot\fett{d}\big)
P_{\mathtt{B}_i}\Theta^i=\mathtt{B}_j\times\Pi$, which immediately
reduces to $-C_{ij}^kP_{\mathtt{B}_k}\Theta^i
=P_{\mathtt{B}_i}C_{jk}^i\Theta^k$.

Infinitesimal equivariance \cite{Marsden} implies that
$P_{\mathtt{A}\times\mathtt{B}}=\{P_{\mathtt{A}},P_{\mathtt{B}}\}_{PB}$.
Then it is easy to see that equivariant momentum maps are Poisson
maps, i.e.\ for scalar-valued functions $F$ and $G$ on
$\mathfrak{g}^*$ one has
\begin{equation}
\{F,G\}_{LPB}(\Pi(\fett{x}))=\{F(\Pi(\fett{x})),G(\Pi(\fett{x}))\}_{PB}.
\label{GHLPBGHPB}
\end{equation}
To prove this one shows that the left hand side of
(\ref{GHLPBGHPB}) can be written as
\begin{equation}
\{F,G\}_{LPB}(\Pi(\fett{x}))=\overline{\mathtt{d}F\times\mathtt{d}G}\cdot
\Pi(\fett{x})=P_{\mathtt{d}F\times\mathtt{d}G}=\{P_{\mathtt{d}G},P_{\mathtt{d}H}\}_{PB},
\end{equation}
where one uses in the last step infinitesimal equivariance. The
right hand side of (\ref{GHLPBGHPB}) gives the same result:
\begin{equation}
\{F(\Pi(\fett{x})),G(\Pi(\fett{x}))\}_{PB}=J^{ij}\partial_i
F(\Pi(\fett{x}))\partial_j G(\Pi(\fett{x}))=J^{ij}\partial_i
P_{\mathtt{d}F}\partial_j P_{\mathtt{d}G}
=\{P_{\mathtt{d}F},P_{\mathtt{d}G}\}_{PB},
\end{equation}
with $\partial_iF(\Pi(\fett{x}))
=\overline{\mathtt{d}F}\cdot\partial_i\Pi(\fett{x})
=\partial_i\left(\overline{\mathtt{d}F}\cdot\Pi(\fett{x})\right)
=\partial_iP_{\mathtt{d}F}$.

A special case for a momentum map is the momentum map of the
cotangent lift of a rotor action on a vector manifold
$\fett{q}=q^a(q^i)\fett{\sigma}_a$. In order to find this momentum
map one first states that it is possible to find for a tangent
vector field $\fett{a}(\fett{q}) =a^i\xi_i^a\fett{\sigma}_a$ a
function $P_{\mbox{\scriptsize{\boldmath$a$}}}(q^i,p_i)
=P_{\mbox{\scriptsize{\boldmath$a$}}}(\fett{q}+\fett{\pi})$ on the
cotangent bundle, which is given with the projection operator
(\ref{Tpiqdef}) as:
\begin{equation}
P_{\mbox{\scriptsize{\boldmath$a$}}}(q^i,p_i)
=T\pi_{\mbox{\scriptsize{\boldmath$q$}}}
(\fett{a})\cdot(\fett{q}+\fett{\pi})=a^j\xi_j^a\fett{\tau}_a
\cdot(q^b\fett{\sigma}_b+p_k\xi^k_b\fett{\tau}^b)=a^j(q^i)p_j.
\end{equation}
These functions form an algebra on the cotangent bundle, i.e.
$\{P_{\mbox{\scriptsize{\boldmath$a$}}},P_{\mbox{\scriptsize{\boldmath$b$}}}\}_{PB}
=-P_{\left[\mbox{\scriptsize{\boldmath$a$}},\mbox{\scriptsize{\boldmath$b$}}\right]_{JLB}}
\phantom{\fett{0}}\!\!\!$. The rotor action of a rotor
$R(t)=e_{\starP}^{\frac{t}{2}\mathtt{B}}$ on the vector manifold
$\fett{q}$ induces a flow $\fett{q}(t)
=R(t)\starP\fett{q}\starP\overline{R(t)}$ and a tangential vector
field $\fett{b}=\mathtt{B}\cdot\fett{q}$. The inverse cotangent
lift of this rotor action is
\begin{equation}
(\fett{q}+\fett{\pi})(t)=\overline{R_{\mathrm{lifted}}(-t)}
\starP(\fett{q}+\fett{\pi})\starP R_{\mathrm{lifted}}(-t)
=R_{\mathrm{lifted}}(t)\starP(\fett{q}+\fett{\pi})\starP
\overline{R_{\mathrm{lifted}}(t)},
\end{equation}
which induces on the cotangent bundle a tangent vector field
$\fett{b}_{\mathrm{lifted}}
=\mathtt{B}_{\mathrm{lifted}}\cdot(\fett{q}+\fett{\pi})$. The
vector field $\fett{b}_{\mathrm{lifted}}$ is then the hamiltonian
vector field of $P_{\mbox{\scriptsize{\boldmath$b$}}}
\phantom{\fett{0}}\!\!\!$, i.e.\ $\fett{b}_{\mathrm{lifted}}
=\fett{h}_{P_{\mbox{\scriptsize{\boldmath$b$}}}}\phantom{\fett{0}}\!\!\!$.
This can be proved very easily if one considers that the cotangent
lift of a rotor action leaves the canonical one-form invariant,
i.e.\
$\mathscr{L}_{\mbox{\scriptsize{\boldmath$b$}}_{\mathrm{lifted}}}
\fett{\theta}=0$. Cartan's magic formula (\ref{magicCartan}) gives
then
\begin{equation}
\fett{b}_{\mathrm{lifted}}\cdot\Omega
=-\dot{\iota}_{\mbox{\scriptsize{\boldmath$b$}}_{\mathrm{lifted}}}
\fett{d}\fett{\theta}=\fett{d}\dot{\iota}
_{\mbox{\scriptsize{\boldmath$b$}}_{\mathrm{lifted}}}\fett{\theta}
=\fett{d}(\fett{b}_{\mathrm{lifted}}\cdot\fett{\theta}).\label{bliftmagic}
\end{equation}
On the other hand one has with (\ref{thetadef}) and
(\ref{Tpiqdef})
\begin{equation}
\fett{b}_{\mathrm{lifted}}\cdot\fett{\theta}(\fett{q}+\fett{\pi})
=T\pi_{\mbox{\scriptsize{\boldmath$q$}}}
(\fett{b}_{\mathrm{lifted}})\cdot\fett{\pi}
=T\pi_{\mbox{\scriptsize{\boldmath$q$}}}(\fett{b})\cdot\fett{\pi}
=P_{\mbox{\scriptsize{\boldmath$b$}}}(\fett{q}+\fett{\pi}).
\end{equation}
Putting this into (\ref{bliftmagic}) gives
\begin{equation}
\fett{b}_{\mathrm{lifted}}\cdot\Omega
=\fett{d}P_{\mbox{\scriptsize{\boldmath$b$}}}\phantom{\fett{0}}\!\!\!,
\end{equation}
which shows that $\fett{b}_{\mathrm{lifted}}$ is the hamiltonian
vector field of $P_{\mbox{\scriptsize{\boldmath$b$}}}
\phantom{\fett{0}}\!\!\!$.

The momentum map of the cotangent lift of a rotor action on the
vector manifold $\fett{q}$ is then given for
$\fett{b}=\mathtt{B}\cdot\fett{q}$ by
\begin{equation}
\overline{\mathtt{B}}\cdot\Pi(\fett{q}+\fett{\pi})
=T\pi_{\mbox{\scriptsize{\boldmath$q$}}}(\mathtt{B}\cdot\fett{q})
\cdot(\fett{q}+\fett{\pi})=P_{\mbox{\scriptsize{\boldmath$b$}}}
(\fett{q}+\fett{\pi}).
\end{equation}
Moreover this momentum map is also equivariant:
\begin{eqnarray}
\overline{\mathtt{B}}\cdot\Pi(R_{\mathrm{lifted}}\starP
(\fett{q}+\fett{\pi})\starP\overline{R_{\mathrm{lifted}}})
&=&T\pi_{\mbox{\scriptsize{\boldmath$q$}}}\phantom{\fett{0}}\!\!\!
\big(\mathtt{B} \cdot(R\starP\fett{q}\starP\overline{R})\big)\cdot
(R_{\mathrm{lifted}}\starP(\fett{q}+\fett{\pi})\starP
\overline{R_{\mathrm{lifted}}})\\
&=&T\pi_{\mbox{\scriptsize{\boldmath$q$}}}
(\mathrm{Ad}_{\overline{R}}\mathtt{B}\cdot\fett{q})\cdot(\fett{q}+\fett{\pi})\\
&=&\overline{\mathrm{Ad}_{\overline{R}}\mathtt{B}}\cdot
\Pi(\fett{q}+\fett{\pi}).
\end{eqnarray}

A simple example is the action of the rotation group on a three
dimensional euclidian vector space with vectors
$\fett{q}=q^i\fett{\eta}_i$ for $i=1,2,3$. The tangent bundle is
then a six dimensional euclidian vector space with vectors
$\fett{q}+\fett{\pi}=q^i\fett{\eta}_i+p_i\fett{\rho}^i$ and a
canonical symplectic structure
$\Omega=\fett{\eta}^i\fett{\rho}_i$. A rotation on the
$\fett{q}$-space is generated by the bivectors $\mathtt{B}_i
=\frac{1}{2}\varepsilon_{ijk}\fett{\eta}_j\fett{\eta}_k$. For
example a rotation around the $\fett{\eta}_3$-axis is generated by
$\mathtt{B}_3=\fett{\eta}_1\fett{\eta}_2$ and the corresponding
vector field is $\fett{b}_3 =\mathtt{B}_3\cdot\fett{q}
=q^2\fett{\eta}_1-q^1\fett{\eta}_2$. The lifted rotation is a
rotation that acts in the $\fett{\rho}_i$-space just the same way
as in the $\fett{\eta}_i$-space, the lifted generator is then
$\mathtt{B}_3^{\mathrm{lifted}}=\fett{\eta}_1\fett{\eta}_2
+\fett{\rho}^1\fett{\rho}^2$ and the corresponding lifted vector
field is given by
\begin{equation}
\fett{b}^{\mathrm{lifted}}_3
=\mathtt{B}^{\mathrm{lifted}}_3\cdot(\fett{q}+\fett{\pi})
=q^2\fett{\eta}_1-q^1\fett{\eta}_2+p_2\fett{\rho}^1-p_1\fett{\rho}^2.
\label{vBindu}
\end{equation}
The Hamilton function $P_{\mathtt{B}_3}$ that generates this
vector field fulfills $\fett{b}^{\mathrm{lifted}}_3\cdot\Omega
=\fett{d}P_{\mathtt{B}_3}$ or
\begin{equation}
p^2\fett{\eta}^1-p^1\fett{\eta}^2-q^2\fett{\rho}^1+q^1\fett{\rho}^2
=\fett{\eta}^1\frac{\partial P_{\mathtt{B}_3}}{\partial q^1}
+\fett{\eta}^2\frac{\partial P_{\mathtt{B}_3}}{\partial q^2}
+\fett{\rho}_1\frac{\partial P_{\mathtt{B}_3}}{\partial p_1}
+\fett{\rho}_2\frac{\partial P_{\mathtt{B}_3}}{\partial p_2},
\end{equation}
which is solved by the angular momentum function. The angular
momentum functions $P_{\mathtt{B}_i}=\varepsilon_{ij}^kq^jp_k$ are
the generators of the active rotations, that rotate the $q^i$ just
as the $p_i$ coefficients. They form the algebra
$\{P_{\mathtt{B}_i},P_{\mathtt{B}_j}\}_{PB}=\varepsilon_{ijk}
P_{\mathtt{B}_k}$, so that there is a hamiltonian action of the
rotations on the six dimensional symplectic space. The momentum
map $\Pi(q^i,p_i)=P_{\mathtt{B}_j}(q^i,p_i)\Theta^j$ is just the
angular momentum bivector $\mathtt{L}=\fett{qp}$ and connects the
generators of the active and passive rotations.

Another simple example is the circle action of $S^1$ on $S^2$
\cite{McDuff}. The two-dimensional sphere
$\fett{x}(\theta,\varphi)=\sin\theta\cos\varphi\fett{\sigma}_1
+\sin\theta\sin\varphi\fett{\sigma}_2+\cos\theta\fett{\sigma}_3$
is a symplectic vector manifold with the symplectic two-form
\begin{equation}
\Omega=x^1\fett{\sigma}^2\fett{\sigma}^3+x^2\fett{\sigma}^3\fett{\sigma}^1
+x^3\fett{\sigma}^1\fett{\sigma}^2\big|_{S^2}=\sin\theta\fett{\xi}^{\theta}
\fett{\xi}^{\varphi},
\end{equation}
which is the volume form on the $S^2$. A left rotation around the
$\fett{\sigma}_3$-axis is generated by $\mathtt{B}
=-\fett{\sigma}_1 \fett{\sigma}_2$ and induces on $S^2$ the vector
field
\begin{equation}
\mathtt{B}\cdot\fett{x}=\sin\theta\cos\varphi\fett{\sigma}_2-
\sin\theta\sin\varphi\fett{\sigma}_1=\partial_{\varphi}\fett{x}
=\fett{\xi}_{\varphi}.
\end{equation}
The Hamilton function $P_{\mathtt{B}}$ that generates this vector
field fulfills according to (\ref{hHOmegadH}) the equation
$\fett{\xi}_{\varphi}\cdot\Omega=\fett{d}P_{\mathtt{B}}$, or
\begin{equation}
-\sin\theta\fett{\xi}^{\theta}=\fett{\xi}^{\varphi}
\partial_{\varphi}P_{\mathtt{B}}
+\fett{\xi}^{\theta}\partial_{\theta}P_{\mathtt{B}},
\end{equation}
which is solved by $P_{\mathtt{B}}=\cos\theta=x^3$.

Applying now the concepts discussed so far to the cotangent space
of a group manifold $T^*G$, which is a vector manifold with
vectors $\fett{r}+\fett{\vartheta}$, one arrives at the
Lie-Poisson reduction \cite{Marsden}. As seen above the rotors act
on the group vector manifold with a left translation $\ell_R$
which induces the tangential maps $T\ell_R$ and $T^*\ell_R$. A
scalar function $F(\fett{r}+\fett{\vartheta})=F(R,\dot{R})$ on
$T^*G$ is left invariant if $F\circ T^*\ell_R=F$. Such left
invariant functions can be identified with reduced functions on
$\mathfrak{g}$, i.e.\ $F(\fett{r}+\fett{\vartheta})=F(R,\dot{R})
=F(1,\overline{R}\starP\dot{R})=f(\Theta)$, where
$\overline{R}\starP\dot{R}$ is an element of the bivector algebra
that can also be expressed in the dual basis. This reduction can
now be described with the momentum map
$\Pi:T^*G\rightarrow\mathfrak{g}^*$, i.e.
$F(\fett{r}+\fett{\vartheta})=f(\Pi(\fett{r}+\fett{\vartheta}))$.
One has then a Poisson map between the Poisson bracket of left
invariant functions on $T^*G$ and the Lie-Poisson bracket of
reduced functions on $\mathfrak{g}^*$. In this way a left
invariant Hamilton function on $T^*G$ induces a Lie-Poisson
dynamic on $\mathfrak{g}^*$. This will be explained for the
example of the rigid body in the next section.

\section{The rigid Body}
\setcounter{equation}{0}
The rigid body is an example where the formalism described above
can be shown to work very effectively. If one considers a free
rigid body $\cal{B}$ in a three-dimensional ambient space spanned
by the basis vectors $\fett{\sigma}_a$ and a body-fixed coordinate
system $\fett{\xi}_i(t)$, a point of the body in the ambient space
is given by
\begin{equation}
\fett{x}(t)=R(t)\starP\fett{x}_{\cal{B}}\starP \overline{R}(t),
\label{xRxBoliR}
\end{equation}
where $\fett{x}_{\cal{B}}$ is the vector in the body-fixed system.
The velocity is then given by
\begin{eqnarray}
\dot{\fett{x}}&=&\dot{R}\starP\fett{x}_{\cal{B}}\starP\overline{R}
+R\starP\fett{x}_{\cal{B}}\starP\dot{\overline{R}}\\
&=&R\starP(\overline{R}\starP\dot{R}\starP\fett{x}_{\cal{B}}
-\fett{x}_{\cal{B}}\starP\overline{R}\starP\dot{R})\starP\overline{R}\\
&=&\dot{R}\starP\overline{R}\starP\fett{x}-\fett{x}\starP\dot{R}\starP
\overline{R}\\
&=&2(\dot{R}\starP\overline{R})\cdot\fett{x},
\end{eqnarray}
using $\overline{R}\starP R=1 \Rightarrow \dot{\overline{R}}\starP
R+\overline{R}\starP\dot{R}=0$. And for the body-fixed velocity
one obtains
\begin{equation}
\dot{\fett{x}}_{\cal{B}}=\overline{R}\starP\dot{\fett{x}}\starP R
=2(\overline{R}\starP\dot{R})\cdot\fett{x}_{\cal{B}}.\label{dotxB}
\end{equation}
On the other side one has $\dot{\fett{x}}=\fett{\omega}
\times\fett{x}$, where $\fett{\omega}$ is the axial vector of
angular velocity. Using that the vector cross product can be
written as $\fett{a}\times\fett{b}=-(I_{(3)}\starP\fett{a})\cdot
\fett{b}$ this leads to
\begin{equation}
\dot{\fett{x}}=-(I_{(3)}\starP\fett{\omega})\cdot\fett{x}
=-\mathtt{W}\cdot\fett{x},\label{x-Wx}
\end{equation}
where
\begin{equation}
\mathtt{W}=-2\dot{R}\starP\overline{R}=I_{(3)}\starP\fett{\omega}
\label{mathttWdef}
\end{equation}
is the angular velocity bivector that generates the rotation.
Equation (\ref{mathttWdef}) can be rewritten to obtain the rotor
equation
\begin{equation}
\dot{R}=-\frac{1}{2}\mathtt{W}\starP R,
\end{equation}
which integrates for constant angular velocity to
$R=e_{\starP}^{-\frac{t}{2}\mathtt{W}}$. With the angular velocity
bivector (\ref{mathttWdef}) one can also write (\ref{x-Wx}) as
\begin{equation}
\dot{\fett{x}}=(R\starP\fett{x}_{\cal{B}}\starP\overline{R})\cdot
\mathtt{W}=R\starP(\fett{x}_{\cal{B}}\cdot\mathtt{W}_{\cal{B}})\starP
\overline{R},
\end{equation}
where $\mathtt{W}_{\cal{B}}=\overline{R}\starP\mathtt{W}\starP R
=-2\overline{R}\starP\dot{R}$, so that the rotor equation becomes
$\dot{R}=-\frac{1}{2}R\starP\mathtt{W}_{\cal{B}}$.

The angular momentum bivector is given by
\begin{eqnarray}
\mathtt{L}&=&\int d^3x\,\rho(\fett{x})\,\fett{x}\dot{\fett{x}}
=\int d^3x_{\cal{B}}\,\rho(\fett{x}_{\cal{B}}) \,
(R\starP\fett{x}_{\cal{B}}\starP\overline{R})
(R\starP(\fett{x}_{\cal{B}}\cdot\mathtt{W}_{\cal{B}})\starP\overline{R})\\
&=&R\starP\left(\int d^3x_{\cal{B}}\,\rho(\fett{x}_{\cal{B}})\,
\fett{x}_{\cal{B}}(\fett{x}_{\cal{B}}\cdot\mathtt{W}_{\cal{B}})\right)
\starP\overline{R}=R\starP\mathtt{I}(\mathtt{W}_{\cal{B}})\starP\overline{R},
\end{eqnarray}
where the bivector-valued function of a bivector
\begin{equation}
\mathtt{I}(\mathtt{B})=\int d^3x_{\cal{B}}\,
\rho(\fett{x}_{\cal{B}}) \,\fett{x}_{\cal{B}}
(\fett{x}_{\cal{B}}\cdot\mathtt{B}),
\end{equation}
corresponds to the inertial tensor. The equation of motion of the
free rigid body can be obtained from
\begin{eqnarray}
0=\dot{\mathtt{L}}
&=&\dot{R}\starP\mathtt{I}(\mathtt{W}_{\cal{B}})\starP\overline{R}
+R\starP\mathtt{I}(\mathtt{W}_{\cal{B}})\starP\dot{\overline{R}}
+R\starP\mathtt{I}(\dot{\mathtt{W}}_{\cal{B}})\starP\overline{R} \label{dotL01}\\
&=&R\starP\big(\mathtt{I}(\dot{\mathtt{W}}_{\cal{B}})
-\mathtt{W}_{\cal{B}}\times\mathtt{I}(\mathtt{W}_{\cal{B}})\big)
\starP\overline{R}\label{dotL02}
\end{eqnarray}
as $\mathtt{I}(\dot{\mathtt{W}}_{\cal{B}})
-\mathtt{W}_{\cal{B}}\times\mathtt{I}(\mathtt{W}_{\cal{B}})=0$,
which are the Euler equations.

The other possibility to derive the equations of motion is to use
the Lagrange or Hamilton formalism. The kinetic energy of the free
rigid body can be written with (\ref{dotxB}) as
\begin{eqnarray}
T&=&\frac{1}{2}\int d^3x_{\cal{B}}\,\rho(\fett{x}_{\cal{B}})\,
|2(\overline{R}\starP\dot{R})\cdot\fett{x}_{\cal{B}}|^2 \label{TRdotR}\\
&=&\frac{1}{2}\int d^3x_{\cal{B}}\,\rho(\fett{x}_{\cal{B}})\,
|\mathtt{W}_{\cal{B}}\cdot\fett{x}_{\cal{B}}|^2\\
&=&\frac{1}{2}\overline{\mathtt{W}_{\cal{B}}}\cdot
\mathtt{I}(\mathtt{W}_{\cal{B}})\label{TmathttWB}\\
&=&\frac{1}{2}\overline{\mathtt{W}}\cdot\mathtt{L}.
\end{eqnarray}
Equation (\ref{TRdotR}) is the left invariant Lagrangian
$L(R,\dot{R})$ and (\ref{TmathttWB}) the reduced Lagrangian
$l(\mathtt{W}_{\cal{B}})$ of the free rigid body. This means that
the dynamics is transferred by (\ref{xRxBoliR}) from the vectors
$\fett{x}(t)$ to the rotors or the generating bivectors, i.e.\ one
considers the dynamics on the rotor group or the bivector algebra
respectively, which is the same idea that underlies the
Kustaanheimo-Stiefel transformation.

The question is now how to vary the corresponding Lagrangians. In
analogy to the matrix representation \cite{Marsden} one has
\begin{eqnarray}
\delta\mathtt{W}_{\cal{B}}=\delta(-2\overline{R}\starP\dot{R})&=&
2\overline{R}\starP\delta R\starP\overline{R}\starP\dot{R}
-2\overline{R}\starP \delta\dot{R}\\
&=&-\overline{R}\starP\delta R\starP \mathtt{W}_{\cal{B}}
-2\overline{R}\starP\delta\dot{R}
\end{eqnarray}
and defining the bivector $\mathtt{B}=2\overline{R}\starP\delta R$
so that
\begin{equation}
\dot{\mathtt{B}}=\mathtt{W}_{\cal{B}}\starP\frac{1}{2}\mathtt{B}+2\overline{R}
\starP\delta\dot{R},
\end{equation}
one obtains
\begin{equation}
\delta\mathtt{W}_{\cal{B}}=-\dot{\mathtt{B}}+\mathtt{W}_{\cal{B}}\times\mathtt{B}.
\label{deltattWB}
\end{equation}

The variation
\begin{eqnarray}
0=\delta l(\mathtt{W}_{\cal{B}})=\delta\int
dt\,\frac{1}{2}\overline{\mathtt{W}_{\cal{B}}}\cdot
\mathtt{I}(\mathtt{W}_{\cal{B}})&=& \int dt\int d^3x_{\cal{B}}\,
\rho(\fett{x}_{\cal{B}})\,
\overline{\delta\mathtt{W}_{\cal{B}}}\cdot\left[\fett{x}_{\cal{B}}
(\fett{x}_{\cal{B}}\cdot\mathtt{W}_{\cal{B}})\right]\label{deltal1}\\
&=&\int dt\, \overline{\mathtt{I}(\mathtt{W}_{\cal{B}})}\cdot
\left(-\dot{\mathtt{B}}+\mathtt{W}_{\cal{B}}\times\mathtt{B}\right)\label{deltal2}\\
&=&\int dt\left[\mathtt{I}(\dot{\mathtt{W}}_{\cal{B}})
+\mathtt{I}(\mathtt{W}_{\cal{B}})\times\mathtt{W}_{\cal{B}}\right]
\cdot\overline{\mathtt{B}},
\end{eqnarray}
leads then again to the Euler equations, where one uses in
(\ref{deltal1})
\begin{equation}
\overline{\mathtt{W}_{\cal{B}}}\cdot\left[\fett{x}_{\cal{B}}
(\fett{x}_{\cal{B}}\cdot\delta\mathtt{W}_{\cal{B}})\right]=
\overline{\delta\mathtt{W}_{\cal{B}}}\cdot\left[\fett{x}_{\cal{B}}
(\fett{x}_{\cal{B}}\cdot\mathtt{W}_{\cal{B}})\right]
\end{equation}
and in (\ref{deltal2}) equation (\ref{deltattWB}).

So given a left invariant rotor Lagrangian $L(R,\dot{R})$ and its
reduction to the bivector algebra $l(\mathtt{W}_{\cal{B}})$, the
variation of $L(R,\dot{R})$ corresponds to the variation of
$l(\mathtt{W}_{\cal{B}})$ for variations
$\delta\mathtt{W}_{\cal{B}}=-\dot{\mathtt{B}}
+\mathtt{W}_{\cal{B}}\times\mathtt{B}$, where $\mathtt{B}$ is a
bivector that vanishes at the endpoints. The Euler-Lagrange
equation for the rotor corresponds to the bivector equation
\begin{equation}
\frac{d}{dt}\frac{\delta l}{\delta\mathtt{W}_{\cal{B}}}
=\mathtt{W}_{\cal{B}}\times \frac{\delta
l}{\delta\mathtt{W}_{\cal{B}}}.
\end{equation}
The Euler-Poincar\'{e} reconstruction of the rotor from the
bivector $\mathtt{W}_{\cal{B}}$ can then be done with the rotor
equation and in a last step the dynamics $\fett{x}(t)$ is
reobtained by (\ref{xRxBoliR}).

In the Hamilton formalism the analogous construction is called
Lie-Poisson reduction and can also be done in the rotor case. The
Hamiltonian (\ref{TmathttWB}) of the free rigid body can be
written as
\begin{equation}
H=\frac{1}{2}\left(\frac{L_{{\cal B}1}^2}{I_1}+\frac{L_{{\cal
B}2}^2}{I_2}+\frac{L_{{\cal B}3}^2}{I_3}\right).\label{rigbodH}
\end{equation}
With the Lie-Poisson bracket (\ref{SO3LPB})
\begin{equation}
\{F,G\}_{LPB}(\mathtt{L}_{\cal B})=\overline{\mathtt{L}_{\cal
B}}\cdot \left((I_{(3)}\starP\fett{\nabla}F)\times (I_{(3)}\starP
\fett{\nabla}G)\right)=\overline{\mathtt{L}_{\cal B}}\cdot
\left(\mathtt{d}F\times\mathtt{d}G\right)
\end{equation}
the Euler equations are obtained by $\dot{L}_{{\cal
B}i}=\{L_{{\cal B}i},H\}_{LPB}$. They preserve the coadjoint
orbit, i.e.\ the Casimir function $|\mathtt{L}_{\cal{B}}|^2$ is a
constant of motion: $\{(L_{{\cal{B}}1}^2+L_{{\cal{B}}2}^2
+L_{{\cal{B}}3}^2),H\}_{LPB} =0$. The conserved quantity that
results from the left invariance is the angular momentum, which
follows from the calculation in (\ref{dotL01}) and (\ref{dotL02}).

The procedure described above is the bivector version of the
Poincar\'{e} equation \cite{Heard}. In order to derive the
Poincar\'{e} equation one considers a vector manifold
$\fett{x}(q^i)$ with coordinate basis vectors
$\fett{\xi}_i=\partial_i\fett{x}$ and non-coordinate basis vectors
$\fett{\vartheta}_r=\vartheta_r^i \fett{\xi}_i$. For a
scalar-valued function $f(q^i(t))$ on a trajectory
$\fett{q}(t)=\fett{x}(q^i(t))$ one has
$\frac{d}{dt}f=\dot{q}^i\partial_if$. In the non-coordinate basis
the coefficients are $s^r=\vartheta^r_i\dot{q}^i$, so that
$\frac{d}{dt}=s^r\partial_r$. On the other hand the variation
of the trajectory $\fett{q}(t)=\fett{q}(t,u=0)$ is given by $\delta
q^i=\left.\frac{d}{du}\right|_{u=0}q^i(t,u)=w^i$, where the
coefficients in the non-coordinate basis are
$w^r=\vartheta_i^rw^i$. From the condition that the operators
\begin{equation}
\frac{d}{dt}=\fett{s}\cdot\fett{\partial}=s^r\partial_r\qquad
\mathrm{and}\qquad \frac{d}{du}=\fett{w}\cdot\fett{\partial}
=w^r\partial_r\label{ddtandddu}
\end{equation}
commute it follows that
\begin{equation}
\frac{d}{du}\fett{s}=\frac{d}{dt}\fett{w}+\left[\fett{s},\fett{w}\right]_{JLB}.
\label{dsduvector}
\end{equation}
This equation can now be used for varying the Lagrange function
$L(q^i(t,u),s^r(t,u))$:
\begin{eqnarray}
0=\delta S&=& \int_a^bdt\left[\frac{\partial L}{\partial
q^i}\frac{\partial q^i}{\partial u}+\frac{\partial L}{\partial
s^r}\left(\frac{d}{dt}w^r+C_{st}^rs^sw^t\right)\right]_{u=0}\\
&=&\int_a^bdt\left[\left(\partial_rL+\frac{\partial L}{\partial
s^s} s^tC_{tr}^s-\frac{d}{dt}\frac{\partial L}{\partial
s^r}\right)w^r+\frac{d}{dt}\left(\frac{\partial L}{\partial s^r}
w^r\right)\right]_{u=0},
\end{eqnarray}
from which the Poincar\'{e} equation follows:
\begin{equation}
\frac{d}{dt}\frac{\partial L}{\partial s^r}-\frac{\partial
L}{\partial s^s}s^tC_{tr}^s=\partial_r L.
\end{equation}

If the configuration space is a rotor group the Lagrange function
is $L=L(R,\dot{R})$ and one has to vary $R(t,u)$. Instead of
vectors $\fett{s}$ and $\fett{w}$ the variations are described by
bivectors
\begin{equation}
\mathtt{s}=2\overline{R}\starP\dot{R}\qquad\mathrm{and}\qquad
\mathtt{w}=2\overline{R}\starP\delta R,
\end{equation}
so that the operators (\ref{ddtandddu}) are now expressed as
$\frac{d}{dt}=\overline{\mathtt{s}}\cdot\mathtt{d}$ and
$\frac{d}{du}=\overline{\mathtt{w}}\cdot\mathtt{d}$. It follows
further that
\begin{eqnarray}
\frac{d\mathtt{s}}{du}&=&-2\overline{R}\starP\delta
R\starP\overline{R}\starP\dot{R}+2\overline{R}\starP\delta\dot{R}
=-\frac{1}{2}\mathtt{w}\starP\mathtt{s}+2\overline{R}\starP\delta\dot{R},\\
\frac{d\mathtt{w}}{dt}&=&-2\overline{R}\starP\dot{R}\starP\overline{R}
\starP\delta R+2\overline{R}\starP\delta\dot{R}
=-\frac{1}{2}\mathtt{s}\starP\mathtt{w}+2\overline{R}\starP\delta\dot{R}.
\end{eqnarray}
Equating the expressions for $2\overline{R}\starP\delta\dot{R}$
gives the bivector analog of (\ref{dsduvector}):
\begin{equation}
\frac{d}{du}\mathtt{s}=\frac{d}{dt}\mathtt{w}+\mathtt{s}\times\mathtt{w},
\end{equation}
so that the variation of the action
\begin{eqnarray}
0=\delta S&=&\int_a^bdt\;\delta L\\
&=&\int_a^bdt\left[\overline{\mathtt{w}}\cdot\mathtt{d}L+\overline{\frac{\delta
L}{\delta\mathtt{s}}}\cdot
\left(\frac{d}{dt}\mathtt{w}+\mathtt{s}\times\mathtt{w}\right)\right]_{u=0}\\
&=&\int_a^bdt\left[\overline{\mathtt{w}}\cdot\left(\mathtt{d}L
-\frac{d}{dt}\frac{\delta L}{\delta\mathtt{s}}+\frac{\delta
L}{\delta\mathtt{s}}\times\mathtt{s}\right)+\frac{d}{dt}\left(
\overline{\mathtt{w}}\cdot\frac{\delta L}{\delta\mathtt{s}}
\right)\right]_{u=0},
\end{eqnarray}
leads now to the bivector version of the Poincar\'{e} equation
\begin{equation}
\frac{d}{dt}\frac{\delta L}{\delta\mathtt{s}}-\frac{\delta
L}{\delta\mathtt{s}}\times\mathtt{s}=\mathtt{d}L.
\end{equation}

In the same way the Hamilton formalism is transferred from the
vector to a bivector basis. The Hamilton equations
$\dot{z}^i=\{z^i,H\}_{PB}$ in the bivector case, i.e. for a
Hamilton function $H(\mathtt{z})$ with a bivector
$\mathtt{z}=z^i\mathtt{B}_i$ are obtained by using the Lie-Poisson
bracket instead of the Poisson bracket. In the
$\mathfrak{so}(3)$-case the Hamilton equations read then
\begin{equation}
\dot{\mathtt{z}}=\mathtt{z}\times\mathtt{d}H
=-\mathrm{ad}^*_{\mathtt{d}H}\mathtt{z}.
\end{equation}

\section*{Conclusions}

\qquad
Comparing classical and quantum mechanics there are two formal
breaks. The first one is that classical mechanics is formulated on
the phase space, while quantum mechanics is formulated on a
Hilbert space. This formal break is overcome by the bosonic star
product formalism that describes quantum mechanics on the phase
space. The second formal break is that classical mechanics is
formulated conventionally in the Gibbs-Heavyside tuple vector
formalism, while in quantum mechanics one is using actually a
Clifford calculus in order to take care of the spin degrees of
freedom. The Gibbs-Heavyside tuple formalism ignores the basis
vectors and their naturally given Clifford structure.
Unfortunately  the basis vectors and their algebraic structure
play an essential role if there is curvature or non-commutativity.
And so the basis vectors had to be reintroduced in the formalism a
posteriori which then naturally leads to a multivector formalism.
The basis vectors appear for example as Dirac matrices, as
differential forms or as Grassmann numbers. These different
formalisms are notationally inconsistent and incomplete. For
example exterior calculus is restricted to homogenous multivectors
and in superanalysis there is no Clifford structure. In the case
of Dirac matrices sticking to a tuple formalism has the
disadvantage that one has to construct an unphysical spinor space
in which the Clifford structure is represented by matrices. A
complete Clifford multivector formalism was on the other hand
developed physically in the context of Dirac theory by Hestenes
and K\"{a}hler and in the context of phase space calculus by Gozzi
and Reuter. The full multivector formalism can now be described
with the star product formalism as deformed superanalysis and so a
formal supersymmetry is introduced in the formalism. The
combination of star products and geometric algebra leads to a
formalism that unifies the different geometric calculi on
commutative and noncommutative spaces, on flat and curved spaces,
on tangent and cotangent spaces and on space-time and phase space.
The combination of the star product formalism with geometric
algebra can be seen as a program for a formal unification of
physics. The consequences of this program on space time and phase
space will be discussed in forthcoming papers. Especially it will
be shown how constraints fit into this context.

\section*{Acknowledgement}

\qquad
The author wants to thank L.\ Schwachh\"ofer for helpful
discussions.


\end{document}